\documentclass[10pt,a4paper]{article}
\pdfoutput=1
\usepackage[english]{babel}
\usepackage[latin1]{inputenc}
\usepackage{amsfonts,amsbsy,bm,euscript,mathrsfs}
\usepackage{amssymb,stmaryrd,faktor}
\usepackage[tbtags]{amsmath}
\usepackage[bookmarks=true,colorlinks=true,linkcolor=black,citecolor=black,urlcolor=black,bookmarksnumbered]{hyperref}
\usepackage[nosort]{cite}

\numberwithin{equation}{section}

\makeatletter
\renewcommand\section{\@startsection {section}{1}{\z@}
{-3.5ex \@plus -1ex \@minus -.2ex}
{2.3ex \@plus.2ex}
{\normalfont\Large\bfseries}}
\renewcommand\subsection{\@startsection{subsection}{2}{\z@}
{-3.25ex\@plus -1ex \@minus -.2ex}
{1.5ex \@plus.2ex}
{\normalfont\large\bfseries}}
\makeatother
\newcommand{\foot}[1]{\footnote{#1\vspace{2pt}}}

\begin{document}

\thispagestyle{empty}

\begin{flushright}\small \tt HU-EP-14/44\end{flushright}

\vspace{2em}

\begin{center}

{\Large \bf Towards a two-parameter $\mathbf{q}$-deformation \\ of AdS$\mathbf{_3 \times S^3 \times M^4}$ superstrings}

\vspace{4em}

{Ben Hoare}

\vspace{2em}

{\em Institut f\"ur Physik und IRIS Adlershof, Humboldt-Universit\"at zu Berlin, \\ Zum Gro\ss en Windkanal 6, 12489, Berlin, Germany.}

\vspace{1em}

\href{mailto:ben.hoare@physik.hu-berlin.de}{\tt ben.hoare@physik.hu-berlin.de}

\end{center}

\vspace{4em}

\begin{abstract}\noindent
We construct a two-parameter deformation of the Metsaev-Tseytlin action for
supercosets with isometry group of the form $\widehat{G} \times \widehat{G}$.
The resulting action is classically integrable and is Poisson-Lie symmetric
suggesting that the symmetry of the model is $q$-deformed,
$\mathcal{U}_{q_{_L}}(\widehat{G}) \times \mathcal{U}_{q_{_R}}(\widehat{G})$.
Focusing on the cases relevant for strings moving in AdS$_3 \times S^3 \times
T^4$ and AdS$_3 \times S^3 \times S^3 \times S^1$, we analyze the corresponding
deformations of the AdS$_3$ and $S^3$ metrics. We also construct a
two-parameter $q$-deformation of the $\mathfrak{u}(1)\inplus
\mathfrak{psu}(1|1)^2 \ltimes \mathfrak{u}(1) \ltimes \mathbb{R}^3$-invariant
R-matrix and closure condition, which underlie the light-cone gauge S-matrix
and dispersion relation of the aforementioned string theories. With the
appropriate identification of parameters, the near-BMN limit of the dispersion
relation is shown to agree with that found from the deformed supercoset sigma
model.
\end{abstract}

\newpage

\tableofcontents

\section{Introduction}\label{secint}

In this article we take the first steps towards constructing a two-parameter
integrable deformation of the AdS$_3 \times S^3 \times T^4$ and AdS$_3 \times
S^3 \times S^3 \times S^1$ superstring theories. These backgrounds have the
feature that their symmetry group takes the form $\widehat G \times \widehat
G$. It is this property that underlies the deformation we consider, which
$q$-deforms the symmetry with an independent parameter for each copy of
$\widehat G$.

Looking for such a deformation is motivated by the two-parameter deformation of
the $S^3$ sigma model of Fateev \cite{Fateev:1996ea}. In \cite{Hoare:2014pna}
it was shown that the former is equivalent to the $SU(2)$ case of Klim\v c\'
ik's bi-Yang-Baxter sigma model \cite{Klimcik:2008eq,Klimcik:2014bta}, and it
is this theory that provides the starting point for the deformation of the
aforementioned string sigma models. The bi-Yang-Baxter sigma model is a
two-parameter integrable deformation of the principal chiral model with
Poisson-Lie symmetry, indicating that the symmetry is $q$-deformed.

To generalize the bi-Yang-Baxter sigma model to a deformation of the
superstring theories, it first needs to be reformulated as a deformation of the
symmetric space coset sigma model. A one-parameter integrable deformation
thereof was formulated in \cite{Delduc:2013fga} for which the global symmetry
is $q$-deformed. In the case that the isometry group of the coset space takes
the form $G \times G$, the model can be gauge-fixed to coincide with the
bi-Yang-Baxter sigma model of \cite{Klimcik:2008eq} with the two deformation
parameters identified. Correspondingly both factors of $G$ are deformed in the
same way.

The one-parameter deformation of the symmetric space coset theory
\cite{Delduc:2013fga} was generalized to a deformation of the Metsaev-Tseytlin
supercoset sigma model \cite{Metsaev:1998it} in \cite{Delduc:2013qra}. For the
AdS$_5 \times S^5$ string background the undeformed supercoset model is
equivalent to the Green-Schwarz string with unfixed $\kappa$-symmetry. The
deformed theory was shown to have many of the properties required to continue
to describe a Green-Schwarz string in a Type IIB supergravity background,
although this remains to be proven. Furthermore in \cite{Delduc:2014kha} it was
confirmed that the $PSU(2,2|4)$ symmetry of the undeformed theory is indeed
$q$-deformed.

It follows that a natural question to ask is whether a two-parameter integrable
deformation of the supercoset sigma model can be found in the case that the
isometry group takes the form $\widehat G \times \widehat G$, such that the
bi-Yang-Baxter sigma model is recovered for bosonic cosets. There are two
models of this type that are of interest in the context of AdS string
backgrounds \cite{Zarembo:2010sg,Wulff:2014kja}, $\widehat G = PSU(1,1|2)$ and
$\widehat G = D(2,1;\alpha)$. The corresponding supercoset theories arise in
particular $\kappa$-symmetry gauge-fixings \cite{Babichenko:2009dk} of the
Green-Schwarz string moving in AdS$_3 \times S^3 \times T^4$
\cite{Giveon:1998ns,Pesando:1998wm} and AdS$_3 \times S^3 \times S^3 \times
S^1$ \cite{Elitzur:1998mm} respectively. In this article we will satisfy
ourselves with constructing the deformation of the supercoset sigma model. To
fully demonstrate the existence of a two-parameter integrable deformation of
the string theories with $q$-deformed symmetry the complete supergravity
background would need to be constructed \cite{Lunin:2014tsa}, and a
$\kappa$-symmetry gauge found such that the corresponding Green-Schwarz action
agrees with the deformed supercoset sigma model.

\

The second approach we will take in this article is to investigate the
deformation of the R-matrices underlying the scattering above the BMN string in
light-cone gauge. After light-cone gauge-fixing the deformed AdS$_5 \times S^5$
model of \cite{Delduc:2013qra}, various tree-level amplitudes describing
scattering above the BMN string \cite{Berenstein:2002jq} were computed in
\cite{Arutyunov:2013ega}. With a certain identification of parameters these
were found to coincide with the expansion of the deformed S-matrix of
\cite{Beisert:2008tw,Hoare:2011wr}. This S-matrix was fixed by demanding
invariance under the $q$-deformation of $\mathfrak{psu}(2|2)^2 \ltimes
\mathbb{R}^3$, the undeformed version of which governs the scattering of
excitations above the BMN string in AdS$_5 \times S^5$
\cite{Beisert:2005tm,Beisert:2006ez,Arutyunov:2006yd}.

For integrable AdS$_3 \times S^3 \times M^4$ string backgrounds, the S-matrix
describing scattering above the BMN string is built out of two $\mathfrak{u}(1)
\inplus \mathfrak{psu}(1|1)^2 \ltimes \mathfrak{u}(1) \ltimes
\mathbb{R}^3$-invariant R-matrices \cite{Borsato:2012ud}, while the dispersion
relations of the scattered excitations follow from closure conditions of the
representations. The R-matrices, supplemented with overall factors unfixed by
symmetry, are combined together in various ways depending on the theory under
consideration and the excitations being scattered
\cite{Borsato:2012ud,Borsato:2013qpa,Borsato:2014exa,Hoare:2013ida,Hoare:2013lja,Lloyd:2014bsa}.
We consider a two-parameter $q$-deformation of this symmetry algebra and
construct the corresponding deformation of the R-matrices. It transpires that
only one of the deformations is a genuine deformation of the algebra, as the
other parameter can be absorbed into the representation. The resulting
R-matrices satisfy braiding unitarity relations, Yang-Baxter equations,
crossing relations, and are matrix unitary for certain reality conditions.
Therefore, they have many of the required properties to describe the scattering
of excitations above the BMN string in the integrable deformed backgrounds.

\

The two constructions in this article, the two-parameter deformation of the
supercoset sigma model and the two-parameter deformation of the R-matrices, are
written in terms of different sets of parameters. From the Poisson-Lie symmetry
of the supercoset theory we can make a semiclassical identification of the
parameters defining the action, with the $q$-parameters governing the
deformation of the symmetry. Assuming these same identifications hold in the
deformation of the R-matrix, as was the case for AdS$_5 \times S^5$
\cite{Arutyunov:2013ega,Delduc:2014kha}, we find that, with a particular
identification of the remaining parameters, the dispersion relation of the
quadratic fluctuations above the BMN vacuum agrees with the expansion of the
dispersion relation following from the closure conditions.

\

Throughout the article we will also compare the two-parameter deformation with
another integrable deformation of strings in AdS$_3 \times S^3 \times M^4$
backgrounds, for which the background is supported by a mix of RR and NSNS flux
\cite{Cagnazzo:2012se}. The corresponding deformations of the $S^3$ sigma model
both appear as limits \cite{Hoare:2014pna} of the four-parameter integrable
theory of \cite{Lukyanov:2012zt}. There are many similar structures and
mechanisms arising in the two constructions and hence it is natural to ask
whether there exists a larger family of integrable deformations of AdS$_3
\times S^3 \times M^4$ superstring theories based on Lukyanov's model
\cite{Lukyanov:2012zt}.

\

The outline of the article is as follows. In section \ref{seccos} we review the
bi-Yang-Baxter sigma model, rewriting the theory as a deformation of the coset
sigma model. This allows for the generalization in section \ref{secsup} to a
two-parameter deformation of the Metsaev-Tseytlin action in the case the
supercoset has isometry $\widehat{G} \times \widehat{G}$. The resulting model's
classical integrability is demonstrated via the existence of a Lax connection.
In section \ref{secmet} we explore the corresponding deformations of the $S^3$
and AdS$_3$ metrics. This is followed in section \ref{secsm} with the
construction of the deformed R-matrices. We conclude with comments and a
discussion of open questions.

\section{\texorpdfstring{$\mathbf{S^3}$}{S3} sigma model}\label{seccos}

We start by reviewing the $S^3$ sigma model and Fateev's two-parameter
deformation thereof \cite{Fateev:1996ea}. In \cite{Hoare:2014pna} the latter
was shown to be equivalent to Klim\v c\' ik's two-parameter bi-Yang-Baxter
sigma model \cite{Klimcik:2008eq,Klimcik:2014bta} for the group $SU(2)$. As
the bi-Yang-Baxter sigma model is written in terms of group- and algebra-valued
fields it is the natural setting for the generalization to the superstring in
section \ref{secsup}.

The $S^3$ sigma model can be written as the principal chiral model for the
group $SU(2)$. The action is given by\,\foot{\label{tension}Note that in all
the action formulae in this article we drop an overall factor of $\frac h2$,
where in the context of string theory the coupling $h$ is proportional to the
string tension. Furthermore, we will largely use light-cone coordinates
normalized as $x^\pm = \frac12(x^0 \pm x^1)$, $\partial_\pm = \partial_0 \pm
\partial_1$.}
\begin{equation}\label{su2pcm}
\mathcal{S} = -\frac{1}{2}\int d^2x \; \operatorname{Tr}[\mathcal J_+ \mathcal J_-] \ ,
\end{equation}
where
\begin{equation}\label{jc}
\mathcal J = g^{-1} d g \in \mathfrak{su}(2) \ ,
\end{equation}
is the left-invariant current for the group-valued field $g \in SU(2)$. For
convenience we assume the fields take values in the defining matrix
representation of $\mathfrak{su}(2)$ or $SU(2)$. The action \eqref{su2pcm} has
a global $SU(2) \times SU(2)$ symmetry corresponding to multiplication of $g$
from the left and right by constant elements of $SU(2)$.

It will be important to understand how the action \eqref{su2pcm} is equivalent
to the symmetric space coset sigma model for
\begin{equation}
\frac{F}{F_0} = \frac{SU(2) \times SU(2)}{SU(2)_{\text{diag}}} \ .
\end{equation}
As this is a symmetric space the algebra $\mathfrak{f} = \mathfrak{su}(2)
\oplus \mathfrak{su}(2)$ admits a $\mathbb{Z}_2$ decomposition
\begin{equation}\label{z2decomposition}
\mathfrak{f} = \mathfrak{f}_0 \oplus \mathfrak{f}_2 \ , \qquad [\mathfrak{f}_i,\mathfrak{f}_j] \subset \mathfrak{f}_{i+j \!\! \mod 4} \ .
\end{equation}
Here the subspace $\mathfrak{f}_0$ is the algebra corresponding to $F_0$, i.e.
it is the diagonal subalgebra of $\mathfrak{su}(2) \oplus \mathfrak{su}(2)$,
and $\mathfrak{f}_2$ is the orthogonal complement of $\mathfrak{f}_0$ in
$\mathfrak{f}$. Using a block-diagonal matrix realization of the product group
$F$ the $\mathbb{Z}_2$ decomposition of $\mathfrak{f}$ can be implemented as
follows
\begin{equation}\begin{split}\label{projections}
\EuScript A = \begin{pmatrix} \mathcal A & 0 \\ 0 & \tilde{\mathcal A} \end{pmatrix} \in \mathfrak{f}\ , \qquad
& P_0\EuScript A = \begin{pmatrix} \mathcal A_0 & 0 \\ 0 & \mathcal A_0 \end{pmatrix} = \frac12 \begin{pmatrix} \mathcal A + \tilde{\mathcal A} & 0 \\ 0 & \tilde{\mathcal A} + \mathcal A \end{pmatrix} \ ,
\\
\mathcal A, \ \tilde{\mathcal A} \in \mathfrak{su}(2) \ , \qquad \qquad
& P_2\EuScript A = \begin{pmatrix} \mathcal A_2 & 0 \\ 0 & -\mathcal A_2 \end{pmatrix} = \frac12 \begin{pmatrix} \mathcal A - \tilde{\mathcal A} & 0 \\ 0 & \tilde{\mathcal A} - \mathcal A \end{pmatrix} \ .
\end{split}\end{equation}
It immediately follows that
\begin{equation}
\operatorname{Tr}[\mathfrak{f}_i\mathfrak{f}_j] = 0 \ , \qquad i+j \neq 0 \mod 4 \ .
\end{equation}

The action is then given by
\begin{equation}\label{su2ssc}
\mathcal{S} = - \int d^2 x \; \operatorname{Tr}[\EuScript J_{+} (P_2\EuScript J_{-})] \ ,
\end{equation}
where $\EuScript J$ is a left-invariant current for the group-valued field $f \in
F$
\begin{align}\nonumber
f = & \begin{pmatrix} g & 0 \\ 0 & \tilde g \end{pmatrix} \in F \ , \qquad \qquad \qquad \qquad \qquad \qquad \qquad \quad \ g, \ \tilde g \in SU(2) \ ,
\\\label{jscr}
\EuScript J = & f^{-1} df = \begin{pmatrix} \mathcal J & 0 \\ 0 & \tilde{\mathcal{J}} \end{pmatrix} = \begin{pmatrix} g^{-1} dg & 0 \\ 0 & \tilde g^{-1} d \tilde g \end{pmatrix} \in \mathfrak{f} \ , \qquad \mathcal J, \ \tilde{\mathcal{J}} \in \mathfrak{su}(2) \ .
\end{align}
As a consequence of the symmetric space's algebraic structure the action
\eqref{su2ssc} has an $SU(2)$ gauge symmetry corresponding to multiplication of
$f$ from the right by a local group element, $f_0 \in F_0$. Under this gauge
symmetry $f$ and $\EuScript J_{0,2}$ transform as
\begin{equation}\label{gauge}
f \to f f_0 \ , \qquad
P_0\EuScript J \to f_0{}^{\!\!-1} (P_0\EuScript J) f_0 + f_0{}^{\!\!-1} d f_0 \ , \qquad
P_2\EuScript J \to f_0{}^{\!\!-1} (P_2\EuScript J) f_0 \ .
\end{equation}
To recover the action \eqref{su2pcm} from \eqref{su2ssc} we note that the
$SU(2)$ gauge symmetry \eqref{gauge} can be used to fix $\tilde g =
\mathbf{1}$, i.e. $\tilde{\mathcal J} = 0$. Then using the projection given in
\eqref{projections} and substituting \eqref{jscr} into \eqref{su2ssc} we indeed
arrive at \eqref{su2pcm}. The action \eqref{su2ssc} also has a global $SU(2)
\times SU(2)$ symmetry corresponding to multiplication of $f$ from the left by
a constant element of $F$.

The equation of motion following from the action \eqref{su2ssc} is given by
\begin{equation}\label{eompre}
\partial_+(P_2 \EuScript J_{-}) + [\EuScript J_+, P_2\EuScript J_{-}] + \partial_- (P_2\EuScript J_{+}) + [\EuScript J_-, P_2\EuScript J_{+}] = 0 \ .
\end{equation}
We also recall that $\EuScript J$ is a left-invariant current and hence it
satisfies the flatness equation
\begin{equation}\label{flatness}
\partial_- \EuScript J_+ - \partial_+ \EuScript J_- + [\EuScript J_-,\EuScript J_+] = 0 \ .
\end{equation}
Projecting \eqref{eompre} and \eqref{flatness} onto $\mathfrak{f}_0$ and
$\mathfrak{f}_2$ we see that they are equivalent to the following equations for
$\mathcal J_0$ and $\mathcal J_2$
\begin{equation}\begin{split}\label{eom}
& \partial_+ \mathcal J_{2-} + [\mathcal J_{0+},\mathcal J_{2-}] + \partial_- \mathcal J_{2+} + [\mathcal J_{0-},\mathcal J_{2+}] = 0 \ ,
\\ & \partial_- \mathcal J_{0+} - \partial_+ \mathcal J_{0-} + [\mathcal J_{0-},\mathcal J_{0+}] + [\mathcal J_{2-},\mathcal J_{2+}] = 0 \ ,
\\ & \partial_- \mathcal J_{2+} + [\mathcal J_{0-},\mathcal J_{2+}] - \partial_+ \mathcal J_{2-} - [\mathcal J_{0+},\mathcal J_{2-}] = 0 \ .
\end{split}\end{equation}
These equations follow from the flatness condition for the following Lax
connection
\begin{equation}
\mathcal{L}_\pm = \mathcal J_{0\pm} + z^{\pm 2} \mathcal J_{2\pm} \ ,
\end{equation}
where $z$ is the spectral parameter. This demonstrates the classical
integrability of this model.

\subsection[Two-parameter deformation of the \texorpdfstring{$S^3$}{S3} sigma model]{Two-parameter deformation of the \texorpdfstring{$\mathbf{S^3}$}{S3} sigma model}\label{secbyb}

In this section we describe the two-parameter Poisson-Lie deformation of the
$S^3$ sigma model, the $SU(2)$ bi-Yang-Baxter sigma model
\cite{Klimcik:2008eq,Klimcik:2014bta}. The model is defined in terms of a
constant antisymmetric solution to the non-split modified classical Yang-Baxter
equation
\begin{equation}\label{mcybe}
[RM,RN] - R([RM,N] + [M,RN]) = [M,N] \ ,
\end{equation}
where $R$ should be thought of as an operator acting on elements $M$, $N$ of an
algebra. One standard solution is to take the operator $R$ to kill elements of
the Cartan subalgebra, multiply positive roots by $-i$ and negative roots by
$i$. We furthermore define the following operator
\begin{equation}
R_g = \operatorname{Ad}_g{}^{\!\!\!\!-1} R \operatorname{Ad}_g \ ,
\end{equation}
where $g$ is an element of the group corresponding to the algebra on which $R$
acts. If $R$ is an antisymmetric solution of \eqref{mcybe}, then $R_g$ is also
an antisymmetric solution.

\

The bi-Yang-Baxter sigma model for $SU(2)$ is given by\,\foot{In
\cite{Hoare:2014pna} it was noted that taking the following solution of the
modified classical Yang-Baxter equation for $\mathfrak{su}(2)$
\begin{equation*}
\bar R(i\sigma_3) = 0 \ , \qquad \bar R(i\sigma_1) = i\sigma_2 \ , \qquad \bar R(i\sigma_2) = -i\sigma_1 \ ,
\end{equation*}
where $\sigma_I$ are the standard Pauli matrices, and defining
\begin{equation*}
M = \frac12 \operatorname{Tr}[g \sigma_3 g^{-1} \sigma_3] \ , \qquad
L_{\pm\,I} = \frac1{2i}\operatorname{Tr}[\partial_\pm g g^{-1}\sigma_I] \ , \qquad
R_{\pm\,I} = \frac1{2i}\operatorname{Tr}[g^{-1}\partial_\pm g\sigma_I] \ ,
\end{equation*}
the action \eqref{su2byb} can be rewritten, up to a total derivative, in the
following way
\begin{equation*}
\mathcal{S} = -\int d^2 x \; \frac1{1+\alpha^2+\beta^2+2\alpha\beta M}\big[\frac12\operatorname{Tr}[g^{-1}\partial_+g g^{-1}\partial_- g] - (\alpha L_{+\,3} + \beta R_{+\,3})(\alpha L_{-\,3}+\beta R_{-\,3})\big] \ .
\end{equation*}
In this form it is clear that setting either $\alpha$ or $\beta$ equal to zero
we find the squashed $S^3$ sigma model of \cite{Cherednik:1981df}.}
\begin{equation}\label{su2byb}
\mathcal{S} = -\frac12 \int d^2 x \; \operatorname{Tr} [\mathcal J_+\frac{1}{1- \alpha \bar R_g - \beta \bar R}\mathcal J_-] \ ,
\end{equation}
where $\mathcal J$ is defined in \eqref{jc}, $\bar R$ is a solution of the
modified classical Yang-Baxter equation for the algebra $\mathfrak{su}(2)$ and
$\alpha$ and $\beta$ are parameters. For $\alpha = \beta = 0$ we recover the
undeformed $SU(2)$ principal chiral model \eqref{su2pcm}.

Introducing an $SU(2)$ gauge symmetry, the action \eqref{su2byb} can be written
as
\begin{equation}\label{su2bybgauge}
\mathcal{S} = -\frac12 \int d^2 x \; \operatorname{Tr} [(\mathcal J_+ - \tilde{\mathcal J}_+) \frac{1}{1-\alpha \bar R_g - \beta \bar R_{\tilde g}}(\mathcal J_- - \tilde{\mathcal J}_-)] \ .
\end{equation}
To recall, $\mathcal J$ and $\tilde{\mathcal J}$ are left-invariant currents
for the $SU(2)$ group-valued fields, $g$ and $\tilde g$
\begin{equation}
\mathcal J = g^{-1} d g \ , \qquad \tilde{\mathcal J} = \tilde g^{-1} d \tilde g \ .
\end{equation}
The action \eqref{su2bybgauge} is then invariant under the following gauge
transformation
\begin{equation}
g \to g g_0 \ , \qquad \tilde g \to \tilde g g_0 \ , \qquad \mathcal J \to g_0{}^{\!\!\!-1} \mathcal J g_0 + g_0{}^{\!\!\!-1} d g_0 \ , \qquad \tilde{\mathcal J} \to g_0{}^{\!\!\!-1} \tilde{\mathcal J} g_0 + g_0{}^{\!\!\!-1} d g_0 \ .
\end{equation}
One can immediately see that this freedom can be used to set $\tilde g =
\mathbf{1}$, i.e. $\tilde{\mathcal{J}} = 0$, and recover \eqref{su2byb}.

In order to generalize to the superstring, and also to compare with the
deformation of \cite{Delduc:2013fga}, we recast the bi-Yang-Baxter sigma model
in the language of the symmetric space coset sigma model \eqref{su2ssc}. Let us
consider the following deformed coset action written in terms of the
group-valued field $f \in SU(2) \times SU(2)$ and a solution $R$ of the
modified classical Yang-Baxter equation \eqref{mcybe} for the algebra
$\mathfrak{su}(2) \oplus \mathfrak{su}(2)$
\begin{equation}\label{su2bybcoset}
\mathcal{S} = - \int d^2 x \; \operatorname{Tr}[\EuScript J_+\big( P_2\frac1{1-I_{\varkappa_{_{L,R}}} R_f P_2} \EuScript J_-\big) ] \ ,
\end{equation}
where
\begin{equation}
I_{\varkappa_{_{L,R}}} = \begin{pmatrix} \varkappa_{_L} \mathbf{1} & 0 \\ 0 & \varkappa_{_R} \mathbf{1} \end{pmatrix} \ .
\end{equation}
If we then write \eqref{su2bybcoset} in terms of $g$, $\tilde g$, $\mathcal{J}$
and $\tilde{\mathcal{J}}$ using \eqref{jscr}, and take $R$ to have the form
\begin{equation}
R = \begin{pmatrix} \bar R & 0 \\ 0 & \pm \bar R\end{pmatrix} \ ,
\end{equation}
then identifying $\varkappa_{_L} = 2\alpha$ and $\varkappa_{_R} = \pm 2\beta$
we recover \eqref{su2bybgauge}. It follows that \eqref{su2bybcoset} is
equivalent to the bi-Yang-Baxter sigma model \eqref{su2byb}. Furthermore, the
form \eqref{su2bybcoset} demonstrates explicitly that if $\varkappa_{_L} =
\varkappa_{_R}$ then we find the deformation of the symmetric space coset sigma
model considered in \cite{Delduc:2013fga}.

\

In the following we will use the following identities and definitions
extensively. First
\begin{equation}
\operatorname{Tr}[M(RN)] = - \operatorname{Tr}[(RM)N] \ , \qquad \operatorname{Tr}[M(P_2N)] = \operatorname{Tr}[(P_2M)N] \ , \qquad M, N \in \mathfrak{su}(2) \ ,
\end{equation}
which follow from the fact that $R$ is an antisymmetric solution of the
modified classical Yang-Baxter equation and the $\mathbb{Z}_2$ automorphism of
the algebra respectively. Second, defining $\Delta = f^{-1}\delta f$, we have
the following variational relations
\begin{equation}\label{variation}
\delta \mathcal O^{-1} = -\mathcal O^{-1} \delta \mathcal O \mathcal O^{-1} \ , \qquad
\delta \EuScript J = d \Delta + [\EuScript J,\Delta] \ , \qquad \delta R_f = [R_f,\operatorname{ad}_\Delta] \ .
\end{equation}
Finally it will be useful to introduce the following operators
\begin{equation}
\mathcal{O}_\pm = 1 \pm I_{\varkappa_{_{L,R}}}R_f P_2 \ .
\end{equation}

\

The action \eqref{su2bybcoset} is invariant under an $SU(2)$ gauge symmetry
acting as in \eqref{gauge}, while the $SU(2) \times SU(2)$ global symmetry of
the undeformed model is broken in the deformed action \eqref{su2bybcoset} (or
equivalently \eqref{su2byb}) to the $U(1) \times U(1)$ subgroup corresponding
to the Cartan elements of $SU(2) \times SU(2)$. The $SU(2) \times SU(2)$
symmetry is Poisson-Lie deformed \cite{Klimcik:2008eq,Klimcik:2014bta}, the
classical predecessor to the $q$-deformation, with different deformation
parameters (depending on $\varkappa_{_L}$ and $\varkappa_{_R}$) for each group
factor. Indeed, based on the results of
\cite{Delduc:2013fga,Klimcik:2008eq,Klimcik:2014bta} it is natural to
conjecture the symmetry of this model (at least semiclassically) is
\begin{equation}\label{symbos}
\mathcal{U}_{q_{_L}}(SU(2)) \times \mathcal{U}_{q_{_R}}(SU(2)) \ , \qquad q_{_L} = \exp(-\frac{\varkappa_{_L}}{h}) \ , \qquad q_{_R} = \exp(-\frac{\varkappa_{_R}}{h}) \ ,
\end{equation}
where $h$ is the overall coupling as defined in footnote \ref{tension}.

Let us briefly demonstrate explicitly the presence of a Poisson-Lie symmetry in
the deformed model. If we consider how the action \eqref{su2bybcoset}
transforms under an infinitesimal multiplication of $f$ from the left
\begin{equation}
f \to f + \epsilon f + \mathcal{O}(\epsilon^2) \ , \qquad \epsilon \in \mathfrak{su}(2) \oplus \mathfrak{su}(2) \ ,
\end{equation}
we find
\begin{equation}\label{variationeps}
\delta_\epsilon S = \int d^2 x \; \operatorname{Tr}[\epsilon(\partial_+\EuScript C_- + \partial_- \EuScript C_+ + I_{\varkappa_{_{L,R}}} ([\EuScript C_-, R \EuScript C_+] + [R\EuScript C_-,\EuScript C_+]))] \ ,
\end{equation}
where
\begin{equation}
\EuScript C_\pm = \operatorname{Ad}_f P_2 \mathcal{O}^{-1}_\pm \EuScript J_\pm \ .
\end{equation}
Therefore, in the undeformed case $\EuScript C$ is the usual conserved current.
The deformation in \eqref{variationeps} then takes the standard Poisson-Lie
form for a $q$-deformed symmetry. Furthermore, considering the restriction of
$\epsilon$ to one or other of the two $\mathfrak{su}(2)$ subalgebras, it is
clear that the deformation of one $\mathfrak{su}(2)$ current just depends on
$\varkappa_{_L}$ and the other on $\varkappa_{_R}$. This motivates the
identification in \eqref{symbos}.

\

To investigate the classical integrability of the model we need to compute the
equations of motion. Varying the action \eqref{su2bybcoset} we find
\begin{equation}\begin{split}\label{e}
& \EuScript E = \partial_+(P_2 \mathcal{O}_-^{-1} \EuScript J_-) + [\EuScript J_+,P_2\mathcal{O}_-^{-1} \EuScript J_-]
+ \partial_-(P_2 \mathcal{O}_+^{-1} \EuScript J_+) + [\EuScript J_-,P_2\mathcal{O}_+^{-1} \EuScript J_+]
\\ & \qquad \qquad \quad + I_{\varkappa_{_{L,R}}}\big([R_f P_2 \mathcal{O}_-^{-1} \EuScript J_-,P_2\mathcal{O}_+^{-1} \EuScript J_+]
+ [P_2 \mathcal{O}_-^{-1} \EuScript J_-,R_fP_2\mathcal{O}_+^{-1} \EuScript J_+]\big) = 0 \ .
\end{split}\end{equation}
Let us also recall that as $\EuScript J$ is a left-invariant current it
satisfies the flatness equation
\begin{equation}\label{z}
\EuScript Z = \partial_- \EuScript J_+ - \partial_+ \EuScript J_- + [\EuScript J_-,\EuScript J_+] = 0 \ .
\end{equation}

We will now demonstrate that these equations follow from a Lax connection.
This was originally shown in \cite{Klimcik:2014bta} for the form of the action
\eqref{su2byb}. Here we will formulate everything in terms of $SU(2) \times
SU(2)$ in order to facilitate the generalization to the supercoset in section
\ref{secsup}. First let us define
\begin{equation}
\EuScript K_\pm = \mathcal{O}_\pm^{-1} \EuScript J_\pm \ , \qquad \EuScript K = \begin{pmatrix} \mathcal K & 0 \\ 0 & \tilde{\mathcal K} \end{pmatrix} \ .
\end{equation}
Equations \eqref{e} and \eqref{z} then translate into the following equations
for $\EuScript K$
\begin{equation}\begin{split}\label{ez}
\EuScript E = & \, \partial_+(P_2\EuScript K_-) + [\EuScript K_+, P_2\EuScript K_-] + \partial_-(P_2 \EuScript K_+) + [\EuScript K_-,P_2\EuScript K_+] = 0 \ ,
\\ \EuScript Z = & \, \partial_- \EuScript K_+ - \partial_+ \EuScript K_- + [\EuScript K_-,\EuScript K_+] + I_{\varkappa_{_{L,R}}}^2 [P_2\EuScript K_-,P_2\EuScript K_+] + I_{\varkappa_{_{L,R}}} R_f \EuScript E = 0 \ .
\end{split}\end{equation}
Projecting these equations onto $\mathfrak{f}_0$ and $\mathfrak{f}_2$ using
\eqref{projections} and defining
\begin{equation}\label{shiftsbos}
\tilde{\mathcal K}_{0} = \mathcal K_{0} + \varkappa_{_+}\varkappa_{_-}\mathcal K_2 \ ,
\qquad \tilde{\mathcal K}_2 = \sqrt{1+\varkappa_{_+}^2}\sqrt{1+\varkappa_{_-}^2}\mathcal K_2 \ ,
\end{equation}
with
\begin{equation}\label{kappmlr}
\varkappa_{_\pm} = \frac12(\varkappa_{_L} \pm \varkappa_{_R}) \ ,
\end{equation}
we find that $\tilde{\mathcal K}_0$ and $\tilde{\mathcal K}_2$ satisfy the
three equations \eqref{eom}. Therefore the Lax connection is given by
\begin{equation}\label{laxkt}
\mathcal{L}_\pm = \tilde{\mathcal K}_{0\pm} + z^{\pm 2} \tilde{\mathcal K}_{2\pm} \ ,
\end{equation}
which in terms of $\mathcal K_{0,2}$ is
\begin{equation}\label{laxk}
\mathcal{L}_\pm = \mathcal K_{0\pm} + \varkappa_{_+}\varkappa_{_-}\mathcal K_{2\pm} + z^{\pm 2} \sqrt{1+\varkappa_{_+}^2}\sqrt{1+\varkappa_{_-}^2}\mathcal K_{2\pm} \ .
\end{equation}
One can then also construct the Lax connection for the original currents
$\mathcal J_{0,2}$.

The necessity of starting from a symmetric space coset sigma model with
symmetry group of the form $G \times G$ is clear from \eqref{shiftsbos}. This
structure allowed us to write the full set of equations given in \eqref{ez} in
terms of $\mathcal{K}_0$ and $\mathcal{K}_2$, which both take values in one
copy of the algebra $\mathfrak{su}(2)$, with no restrictions. Consequently we
could shift one by the other in \eqref{shiftsbos}.

\subsection[\texorpdfstring{$S^3$}{S3} with B-field]{\texorpdfstring{$\mathbf{S^3}$}{S3} with B-field}\label{secmf}

Let us briefly recall that introducing a B-field to the $S^3$ sigma model is
also a deformation that preserves integrability. As a deformation of the
principal chiral model \eqref{su2pcm} the action is given by
\begin{equation}\label{su2pcmmf}
\mathcal{S} = -\frac12\int d^2x \; \operatorname{Tr}[\mathcal J_+ \mathcal J_-] + \frac{\text{b}}3 \int d^3 x \; \epsilon^{abc}\,\operatorname{Tr}[\mathcal J_a\mathcal J_b \mathcal J_c] \ ,
\end{equation}
where $\text{b}$ is a parameter controlling the strength of the B-field. In
particular, $\text{b} = 0$ is the original $SU(2)$ principal chiral model,
while $\text{b}=1$ is the $SU(2)$ WZW model \cite{Witten:1983ar}.

Using the group-valued field $f \in SU(2) \times SU(2)$ introduced in
\eqref{jscr}, the action \eqref{su2pcmmf} can be rewritten as a deformation of
the symmetric space sigma model \eqref{su2ssc} \cite{Cagnazzo:2012se}
\begin{equation}\label{su2sscmf}
\mathcal{S} = - \int d^2x \; \operatorname{Tr}[\EuScript{J}_{+}(P_2\EuScript{J}_{-})] +
\frac{4\text{b}}3\int d^3 x \; \epsilon^{abc} \, \widetilde{\operatorname{Tr}} [(P_2\EuScript{J}_{a})(P_2 \EuScript{J}_{b})(P_2 \EuScript{J}_{c})] \ ,
\end{equation}
where $\widetilde{\operatorname{Tr}}$ is defined as
\begin{equation}
\widetilde{\operatorname{Tr}}[\EuScript A] = \widetilde{\operatorname{Tr}} \begin{pmatrix} \mathcal{A} & 0 \\ 0 & \tilde{\mathcal{A}} \end{pmatrix}
= \operatorname{Tr}[\mathcal{A}] - \operatorname{Tr}[\tilde{\mathcal{A}}] = \operatorname{Tr}[W\EuScript A] \ , \qquad
W = \begin{pmatrix} \mathbf{1} & 0 \\ 0 &-\mathbf{1} \end{pmatrix} \ .
\end{equation}
If the usual trace is used in the WZ term it vanishes as a consequence of the
$\mathbb{Z}_2$ automorphism of the algebra. The action \eqref{su2sscmf} still
has the $SU(2)$ gauge symmetry defined in \eqref{gauge}, which can be used to
fix $\tilde g = \mathbf{1}$, i.e. $\tilde{\mathcal{J}} = 0$, and recover
\eqref{su2pcmmf}. Note that, unlike the two-parameter deformation discussed in
section \ref{secbyb}, the addition of a B-field preserves the global $SU(2)
\times SU(2)$ symmetry of the undeformed model.

The equation of motion following from \eqref{su2sscmf} is
\begin{equation}\label{eomp}
\partial_+ (P_2 \EuScript J_-) + [\EuScript J_+,P_2 \EuScript J_-]
+ \partial_- (P_2 \EuScript J_+) + [\EuScript J_-,P_2 \EuScript J_+] -2 \text{b}W[P_2\EuScript J_-,P_2\EuScript J_+] = 0 \ .
\end{equation}
Projecting this equation and the flatness equation for the left-invariant
current $\EuScript{J}$ \eqref{flatness} onto $\mathfrak{f}_0$ and
$\mathfrak{f}_2$ and defining
\begin{equation}
\tilde{\mathcal{K}}_{0\pm} = \mathcal{J}_{0\pm} \pm \text{b} \mathcal{J}_{2\pm} \ , \qquad \tilde{\mathcal{K}}_{2\pm} = \sqrt{1-\text{b}^2} \mathcal{J}_{2\pm} \ ,
\end{equation}
we find that $\tilde{\mathcal K}_0$ and $\tilde{\mathcal K}_2$ satisfy the
three equations \eqref{eom}, and hence the Lax connection is given by
\begin{equation}
\mathcal{L}_\pm = \tilde{\mathcal K}_{0\pm} + z^{\pm 2} \tilde{\mathcal K}_{2\pm} \ .
\end{equation}
In terms of the original currents $\mathcal J_{0,2}$ the Lax connection is
\begin{equation}\label{laxmixed}
\mathcal{L}_\pm = \mathcal{J}_{0\pm} \pm \text{b} \mathcal{J}_{2\pm} + z^{\pm 2}\sqrt{1-\text{b}^2} \, \mathcal{J}_{2\pm} \ .
\end{equation}
Let us note the similarity with the Lax connection for the two-parameter
deformation written in terms of $\mathcal K_{0,2}$ as given in \eqref{laxk}. In
both cases the part proportional to $z^{\pm 2}$ is rescaled, while the part
proportional to $z^0$ is shifted, with the two light-cone currents shifted in
the same direction for the two-parameter deformation and in opposite directions
for the $S^3$ sigma model with B-field.

The form of the two Lax connections, \eqref{laxk} and \eqref{laxmixed},
suggests that it may be possible to incorporate the two deformations into a
three-parameter deformed model preserving integrability, and indeed such a
theory was constructed in \cite{Lukyanov:2012zt} (see \cite{Hoare:2014pna} for
an explicit demonstration that the four-parameter model of
\cite{Lukyanov:2012zt} has both Fateev's model and the $S^3$ sigma model with a
B-field as limits).

\section{AdS\texorpdfstring{$\mathbf{_3 \times S^3 (\times S^3)}$}{3 x S3 (x S3)} supercoset sigma model}\label{secsup}

We now generalize the bosonic construction described in section \ref{seccos} to
the supercoset case. The supercosets we consider take the form
\begin{equation}\label{supercosets}
\frac{\widehat{F}}{F_0} = \frac{\widehat G \times \widehat G}{F_0} \ ,
\end{equation}
where $F_{0}$ is the bosonic diagonal subgroup of the product supergroup
$\widehat F = \widehat G \times \widehat G$. The supergroup $\widehat G =
PSU(1,1|2)$ is of interest in the context of strings moving in AdS$_3 \times
S^3 \times T^4$ and accordingly has bosonic subgroup $[SU(1,1) \times
SU(2)]^2$. The supergroup $\widehat G =D(2,1;\alpha)$ is relevant for strings
moving in AdS$_3 \times S^3 \times S^3 \times S^1$, with the parameter $\alpha$
related to the radii of the two three-spheres, and as such has bosonic subgroup
$[SU(1,1) \times SU(2) \times SU(2)]^2$.

In the following we will not strictly be talking about the superstring theories
as we will not treat the flat ($T^4$ and $S^1$) directions. In the undeformed
case it is known that there is a (full) $\kappa$-symmetry gauge-fixing
\cite{Babichenko:2009dk} that reduces the Type IIB Green-Schwarz action to the
Metsaev-Tseytlin supercoset action \cite{Metsaev:1998it} plus the requisite
free bosons corresponding to the flat directions. We will describe how to
deform these Metsaev-Tseytlin supercoset actions under the assumption that they
are still $\kappa$-symmetry gauge fixings of consistent 10-dimensional string
theories.

The superalgebra $\hat{\mathfrak{f}}$ corresponding to the product supergroup
$\widehat F$ admits a $\mathbb{Z}_4$ decomposition (the analogue of the
$\mathbb{Z}_2$ decomposition \eqref{z2decomposition} in the bosonic case)
\begin{equation}\label{z4decompositon}
\hat{\mathfrak{f}} = \mathfrak{f}_0 \oplus \mathfrak{f}_1 \oplus \mathfrak{f}_2 \oplus \mathfrak{f}_3 \ , \qquad [\mathfrak{f}_i,\mathfrak{f}_j] \subset \mathfrak{f}_{i+j \!\! \mod 4} \ .
\end{equation}
Here the subspace $\mathfrak{f}_0$ is the algebra corresponding to $F_0$, i.e.
it is the bosonic diagonal subalgebra of $\hat{\mathfrak{f}}$.
$\mathfrak{f}_2$ is the Grassmann-even part of the orthogonal complement of
$\mathfrak{f}_0$ in $\mathfrak{f}$, while $\mathfrak{f}_1$ and $\mathfrak{f}_3$
are the Grassmann-odd parts. We denote the superalgebra corresponding to the
supergroup $\widehat G$ as $\hat{\mathfrak{g}}$. Using a block-diagonal matrix
realization of the product supergroup $\widehat F$ the $\mathbb{Z}_4$
decomposition of $\hat{\mathfrak{f}}$ can be implemented as follows
\begin{equation}\begin{split}\label{projectionsodd}
\EuScript A = \begin{pmatrix} \mathcal A & 0 \\ 0 & \tilde{\mathcal A} \end{pmatrix} \in \hat{\mathfrak{f}} \ , \qquad
& P_0\EuScript A = \begin{pmatrix} \mathcal A_0 & 0 \\ 0 & \mathcal A_0 \end{pmatrix} = \frac12 \begin{pmatrix} P_e(\mathcal A + \tilde{\mathcal A}) & 0 \\ 0 & P_e(\tilde{\mathcal A} + \mathcal A) \end{pmatrix} \ ,
\\
\mathcal A, \ \tilde{\mathcal A} \in \hat{\mathfrak g} \ , \qquad \qquad \quad \ \
& P_1\EuScript A = \begin{pmatrix} \mathcal A_1 & 0 \\ 0 & -i \mathcal A_1 \end{pmatrix} = \frac12 \begin{pmatrix} P_o(\mathcal A + i \tilde{\mathcal A}) & 0 \\ 0 & P_o(\tilde{\mathcal A} - i \mathcal A) \end{pmatrix} \ ,
\\
& P_2\EuScript A = \begin{pmatrix} \mathcal A_2 & 0 \\ 0 & -\mathcal A_2 \end{pmatrix} = \frac12 \begin{pmatrix} P_e(\mathcal A - \tilde{\mathcal A}) & 0 \\ 0 & P_e(\tilde{\mathcal A} - \mathcal A) \end{pmatrix} \ ,
\\
& P_3\EuScript A = \begin{pmatrix} \mathcal A_3 & 0 \\ 0 & i \mathcal A_3 \end{pmatrix} = \frac12 \begin{pmatrix} P_o(\mathcal A -i \tilde{\mathcal A}) & 0 \\ 0 & P_o(\tilde{\mathcal A} + i \mathcal A) \end{pmatrix} \ ,
\end{split}\end{equation}
where $P_e$ and $P_o$ are projections onto the Grassmann-even and Grassmann-odd
parts of the superalgebra $\hat{\mathfrak{g}}$. Defining the supertrace for
$\hat{\mathfrak{g}}\oplus \hat{\mathfrak{g}}$ as the sum of the two supertraces
for each copy of $\hat{\mathfrak{g}}$
\begin{equation}
\operatorname{STr}[\EuScript A] = \operatorname{STr} \begin{pmatrix} \mathcal{A} & 0 \\ 0 & \tilde{\mathcal{A}} \end{pmatrix} = \operatorname{STr}[\mathcal{A}] + \operatorname{STr}[\tilde{\mathcal{A}}] \ ,
\end{equation}
we find immediately that
\begin{equation}
\operatorname{STr}[\mathfrak{f}_i \mathfrak{f}_j] = 0\ , \qquad i+j \neq 0 \mod 4 \ ,
\end{equation}
where we have used the property that the supertrace of the product of an odd
and an even element of the superalgebra is vanishing.

The Metsaev-Tseytlin supercoset action in conformal gauge
\cite{Metsaev:1998it,Berkovits:1999zq} is then given by
\begin{equation}\label{mt}
\mathcal{S} = \int d^2x \; \operatorname{STr}[\EuScript J_+ (P_- \EuScript J_-)]
= \int d^2x \; \operatorname{STr}[(P_+ \EuScript J_+) \EuScript J_-] \ ,
\end{equation}
which we have written in the form appropriate for the deformation
\cite{Delduc:2013qra,Delduc:2014kha}. Here $\EuScript J$ is a left-invariant
current for the supergroup-valued field $f \in \hat F$
\begin{align}\nonumber
f = & \begin{pmatrix} g & 0 \\ 0 & \tilde g \end{pmatrix} \in \widehat F \ , \qquad \qquad \qquad \qquad \qquad \qquad \qquad \quad \ g, \ \tilde g \in \widehat G \ ,
\\
\EuScript J = & f^{-1} df = \begin{pmatrix} \mathcal J & 0 \\ 0 & \tilde{\mathcal{J}} \end{pmatrix} = \begin{pmatrix} g^{-1} dg & 0 \\ 0 & \tilde g^{-1} d \tilde g \end{pmatrix} \in \hat{\mathfrak{f}} \ , \qquad \mathcal J, \ \tilde{\mathcal{J}} \in \hat{\mathfrak{g}} \ ,
\end{align}
while $P_\pm$ are certain linear combinations of the projectors $P_{1,2,3}$
\begin{equation}
P_\pm = P_2 \mp \frac12 (P_1 - P_3) \ .
\end{equation}

As $P_\pm$ do not include $P_0$ the action \eqref{mt} has a $F_0$ gauge
symmetry, which acts as
\begin{equation}\label{gaugesup}
f \to f f_0 \ , \qquad
P_0\EuScript J \to f_0{}^{\!\!-1} (P_0\EuScript J) f_0 + f_0{}^{\!\!-1} d f_0 \ , \qquad
P_{1,2,3}\EuScript J \to f_0{}^{\!\!-1} (P_{1,2,3}\EuScript J) f_0 \ .
\end{equation}
The action \eqref{mt} also has a global $\widehat G \times \widehat G$ symmetry
corresponding to multiplication of $f$ from the left by a constant element of
$\widehat F$.

The equation of motion following from \eqref{mt} and flatness equation for
$\EuScript J$ are given by
\begin{equation}\begin{split}
& \partial_+ (P_-\EuScript J_-) + [\EuScript J_+, P_-\EuScript J_-] + \partial_- (P_+\EuScript J_+) + [\EuScript J_-, P_+\EuScript J_+] = 0 \ ,
\\
& \partial_- \EuScript J_+ - \partial_+ \EuScript J_- + [\EuScript J_-, \EuScript J_+] = 0 \ .
\end{split}\end{equation}
As in the bosonic case we can use \eqref{projectionsodd} to decompose these two
equations, and write them in terms of the $\hat{\mathfrak{g}}$-valued currents
$\mathcal{J}_0$, $\mathcal{J}_1$, $\mathcal{J}_2$ and $\mathcal{J}_3$. Doing so
we find
\begin{equation}\begin{split}\label{fulleom}
& \partial_- \mathcal{J}_{0+} - \partial_+ \mathcal{J}_{0-}
+ [\mathcal{J}_{0-},\mathcal{J}_{0+}]+ [\mathcal{J}_{1-},\mathcal{J}_{3+}]
+ [\mathcal{J}_{2-},\mathcal{J}_{2+}]+ [\mathcal{J}_{3-},\mathcal{J}_{1+}] = 0 \ ,
\\
& \partial_- \mathcal{J}_{2+} + [\mathcal{J}_{0-},\mathcal{J}_{2+}] + [\mathcal{J}_{3-},\mathcal{J}_{3+}] = 0 \ ,
\qquad \partial_+ \mathcal{J}_{2-} + [\mathcal{J}_{0+},\mathcal{J}_{2-}] + [\mathcal{J}_{1+},\mathcal{J}_{1-}] = 0 \ ,
\\
& [\mathcal{J}_{1+},\mathcal{J}_{2-}] = 0 \ , \qquad
\partial_- \mathcal{J}_{1+} + [\mathcal{J}_{0-},\mathcal{J}_{1+}] - \partial_+ \mathcal{J}_{1-} - [\mathcal{J}_{0+},\mathcal{J}_{1-}] +
[\mathcal{J}_{2-},\mathcal{J}_{3+}] = 0\ ,
\\
& [\mathcal{J}_{3-},\mathcal{J}_{2+}] = 0 \ , \qquad
\partial_- \mathcal{J}_{3+} + [\mathcal{J}_{0-},\mathcal{J}_{3+}] - \partial_+ \mathcal{J}_{3-} - [\mathcal{J}_{0+},\mathcal{J}_{3-}] -
[\mathcal{J}_{2+},\mathcal{J}_{1-}] = 0\ .
\end{split}\end{equation}
These equations follow from the flatness condition for the following Lax
connection \cite{Bena:2003wd}
\begin{equation}
\mathcal{L}_\pm= \mathcal J_{0\pm} + z^{-1} \mathcal J_{1\pm} + z \mathcal{J}_{3\pm} + z^{\pm 2} \mathcal J_{2\pm} \ ,
\end{equation}
where $z$ is the spectral parameter. This demonstrates the classical
integrability of this model. The Lax connection is also invariant under the
following $\mathbb{Z}_4$ symmetry
\begin{equation}\label{z4lax}
\mathcal{J}_{k} \to i^k \mathcal{J}_k \ , \qquad z \to i z \ .
\end{equation}

\subsection[Two-parameter deformation of the AdS\texorpdfstring{$_3 \times S^3 (\times S^3)$}{3 x S3 (x S3)} sigma model]{Two-parameter deformation of the AdS\texorpdfstring{$\mathbf{_3 \times S^3 (\times S^3)}$}{3 x S3 (x S3)} sigma model}\label{secsud}

Motivated by the results of \cite{Delduc:2013qra,Delduc:2014kha} a natural
conjecture for the two-parameter deformation of the supercoset action
\eqref{mt} is
\begin{equation}\label{symbybcoset}
\mathcal{S} = \int d^2 x \; \operatorname{STr}[\EuScript J_+\big( P^{\eta_{_{L,R}}}_-\frac1{1-I_{\eta_{_{L,R}}} R_f P^{\eta_{_{L,R}}}_-} \EuScript J_-\big) ] \ ,
\end{equation}
where
\begin{equation}
I_{\eta_{_{L,R}}} = \frac2{\sqrt{1-\eta_{_{L}}^2}\sqrt{1-\eta_{_{R}}^2}} \begin{pmatrix} \eta_{_L} \mathbf{1} & 0 \\ 0 & \eta_{_R} \mathbf{1} \end{pmatrix} \ , \qquad P^{\eta_{_{L,R}}}_\pm = P_2 \mp \frac{\sqrt{1-\eta_{_L}^2}\sqrt{1-\eta_{_R}^2}}2(P_1-P_3) \ ,
\end{equation}
and $R_f$ is defined in terms of an antisymmetric constant solution $R$ of the
modified classical Yang-Baxter equation \eqref{mcybe} for the superalgebra
$\hat{\mathfrak{f}}$ and the supergroup-valued field $f \in \widehat{F}$
\begin{equation}
R_f = \operatorname{Ad}_f^{-1} R \operatorname{Ad}_f \ .
\end{equation}
It is clear from \eqref{symbybcoset} that if $\eta_{_L}= \eta_{_R} = \eta$ we
find the deformation constructed in \cite{Delduc:2013qra,Delduc:2014kha}, while
if we set $\eta_{_L} = \eta_{_R} = 0$ we recover the undeformed model
\eqref{mt}. The normalization of \eqref{symbybcoset} is fixed so that if we
truncate to a bosonic $SU(2)$ sector we recover the action \eqref{su2bybcoset}
with the identification
\begin{equation}\label{abeta}
\varkappa_{_L} = \frac{2\eta_{_L}}{\sqrt{1-\eta_{_L}^2}\sqrt{1-\eta_{_R}^2}} \ , \qquad
\varkappa_{_R} = \frac{2\eta_{_R}}{\sqrt{1-\eta_{_L}^2}\sqrt{1-\eta_{_R}^2}} \ .
\end{equation}

\

In the following we will use the identities
\begin{equation}
\operatorname{STr}[M(RN)] = - \operatorname{STr}[(RM)N] \ , \qquad \operatorname{STr}[M(P^{\eta_{_{L,R}}}_- N)] = \operatorname{STr}[(P^{\eta_{_{L,R}}}_+M)N] \ ,
\end{equation}
which follow from the fact that $R$ is an antisymmetric solution of the
modified classical Yang-Baxter equation and the $\mathbb{Z}_4$ automorphism of
the algebra respectively. It will also be useful to define the operators
\begin{equation}
\mathcal{O}_\pm = 1 \pm I_{\eta_{_{L,R}}} R_f P^{\eta_{_{L,R}}}_\pm \ ,
\end{equation}
and recall the variational relations \eqref{variation}.

\

As in the undeformed case the action \eqref{symbybcoset} is invariant under the
$F_0$ gauge symmetry \eqref{gaugesup}, while the $\widehat{F} =
\widehat{G}\times \widehat{G}$ global symmetry is broken to its Cartan
subgroup. As for the bosonic case, and by analogy with the deformation of the
AdS$_5 \times S^5$ supercoset \cite{Delduc:2014kha}, it is expected that this
symmetry is Poisson-Lie deformed, the classical predecessor to the
$q$-deformation, with different deformation parameters (depending on
$\eta_{_L}$ and $\eta_{_R}$) for each group factor. Indeed, based on the
results of \cite{Delduc:2014kha} it is natural to conjecture the symmetry of
this model (at least semiclassically) is
\begin{equation}\label{symsym}
\mathcal{U}_{q_{_L}}(\widehat G) \times \mathcal{U}_{q_{_R}}(\widehat G) \ , \qquad q_{_L} = \exp(-\frac{\varkappa_{_L}(\eta_{_L},\eta_{_R})}{h}) \ , \qquad q_{_R} = \exp(-\frac{\varkappa_{_R}(\eta_{_L},\eta_{_R})}{h}) \ ,
\end{equation}
where $h$ is an overall coupling (as defined in footnote \ref{tension}), and
$\varkappa_{_L}$ and $\varkappa_{_R}$ are defined in terms of $\eta_{_L}$ and
$\eta_{_R}$ in \eqref{abeta}.

To explicitly see the presence of the Poisson-Lie symmetry in the deformed
model let us consider how the action \eqref{symbybcoset} transforms under an
infinitesimal multiplication of $f$ from the left
\begin{equation}
f \to f + \epsilon f + \mathcal{O}(\epsilon^2)\ , \qquad \epsilon \in \hat{\mathfrak{g}} \oplus \hat{\mathfrak{g}} \ .
\end{equation}
Doing so we find
\begin{equation}\label{variationeps2}
\delta_\epsilon S = - \int d^2 x \; \operatorname{STr}[\epsilon(\partial_+\EuScript C_- + \partial_- \EuScript C_+ + I_{\eta_{_{L,R}}} ([\EuScript C_-, R \EuScript C_+] + [R\EuScript C_-,\EuScript C_+]))] \ ,
\end{equation}
where
\begin{equation}
\EuScript C_\pm = \operatorname{Ad}_f P^{\eta_{_{L,R}}}_\pm \mathcal{O}^{-1}_\pm \EuScript J_\pm \ .
\end{equation}
Therefore, in the undeformed case $\EuScript C$ is the usual conserved current.
The deformation in \eqref{variationeps2} then takes the standard Poisson-Lie
form for a $q$-deformed symmetry. Furthermore, considering the restriction of
$\epsilon$ to one or other of the two $\hat{\mathfrak{g}}$ subalgebras, it is
clear that the deformation of one $\hat{\mathfrak{g}}$ current just depends on
$\varkappa_{_L}$ and the other on $\varkappa_{_R}$ as defined in \eqref{abeta}.
This motivates the identification in \eqref{symsym}.

\

Varying the action \eqref{symbybcoset} we find the following equation of motion
\begin{equation}\begin{split}\label{esup}
& \EuScript E = \partial_+(P^{\eta_{_{L,R}}}_- \mathcal{O}_-^{-1} \EuScript J_-) + [\EuScript J_+,P^{\eta_{_{L,R}}}_-\mathcal{O}_-^{-1} \EuScript J_-]
+ \partial_-(P^{\eta_{_{L,R}}}_+ \mathcal{O}_+^{-1} \EuScript J_+) + [\EuScript J_-,P^{\eta_{_{L,R}}}_+\mathcal{O}_+^{-1} \EuScript J_+]
\\
& \qquad \qquad \quad + I_{\eta_{_{L,R}}}\big([R_f P^{\eta_{_{L,R}}}_- \mathcal{O}_-^{-1} \EuScript J_-,P^{\eta_{_{L,R}}}_+\mathcal{O}_+^{-1} \EuScript J_+]
+ [P^{\eta_{_{L,R}}}_- \mathcal{O}_-^{-1} \EuScript J_-,R_f P^{\eta_{_{L,R}}}_+\mathcal{O}_+^{-1} \EuScript J_+]\big) = 0 \ .
\end{split}\end{equation}
Let us also recall that as $\EuScript J$ is a left-invariant current it
satisfies the flatness equation
\begin{equation}\label{zsup}
\EuScript Z = \partial_- \EuScript J_+ - \partial_+ \EuScript J_- + [\EuScript J_-,\EuScript J_+] = 0 \ .
\end{equation}

Following the construction for the bosonic model in section \ref{secbyb} we
again define
\begin{equation}\label{326}
\EuScript K_\pm = \mathcal{O}_\pm^{-1} \EuScript J_\pm \ , \qquad \EuScript K = \begin{pmatrix} \mathcal K & 0 \\ 0 & \tilde{\mathcal K} \end{pmatrix} \ .
\end{equation}
Equations \eqref{esup} and \eqref{zsup} then translate into the following
equations for $\EuScript K$
\begin{equation}\begin{split}\label{ezsup}
\EuScript E = & \, \partial_+(P^{\eta_{_{L,R}}}_-\EuScript K_-) + [\EuScript K_+, P^{\eta_{_{L,R}}}_-\EuScript K_-] + \partial_-(P^{\eta_{_{L,R}}}_+ \EuScript K_+) + [\EuScript K_-,P^{\eta_{_{L,R}}}_+\EuScript K_+] = 0 \ ,
\\ \EuScript Z = & \, \partial_- \EuScript K_+ - \partial_+ \EuScript K_- + [\EuScript K_-,\EuScript K_+] + I_{\eta_{_{L,R}}}^2 [P^{\eta_{_{L,R}}}_-\EuScript K_-,P^{\eta_{_{L,R}}}_+\EuScript K_+] + I_{\eta_{_{L,R}}} R_f \EuScript E = 0 \ .
\end{split}\end{equation}

Projecting these equations onto $\mathfrak{f}_0$, $\mathfrak{f}_1$,
$\mathfrak{f}_2$ and $\mathfrak{f}_3$ using \eqref{projectionsodd} and defining
\begin{alignat}{2}\nonumber
\tilde{\mathcal K}_{0\pm} = & \mathcal{K}_{0\pm} + \bar \eta \mathcal{K}_{2\pm} \ , \qquad
& \tilde{\mathcal K}_{1\pm} = & \widehat \eta^{\frac12}(\eta_+\mathcal{K}_{1\pm} - \eta_- \mathcal{K}_{3\pm}) \ ,
\\\label{shiftsfull}
\tilde{\mathcal K}_{2\pm} = & \widehat \eta {\mathcal{K}}_{2\pm} \ , \qquad
& \tilde{\mathcal K}_{3\pm} = & \widehat \eta^{\frac12}(\eta_+\mathcal{K}_{3\pm} - \eta_- {\mathcal{K}}_{1\pm}) \ ,
\end{alignat}
with
\begin{equation}
\bar\eta = \frac{\eta_{_L}^2-\eta_{_R}^2}{(1-\eta_{_L}^2)(1-\eta_{_R}^2)} \ , \qquad
\widehat\eta = \frac{1-\eta_{_L}^2\eta_{_R}^2}{(1-\eta_{_L}^2)(1-\eta_{_R}^2)} \ , \qquad
\eta_\pm = \frac{\sqrt{1-\eta_{_L}^2} \pm\sqrt{1-\eta_{_R}^2}}{2} \ ,
\end{equation}
we find that $\tilde{\mathcal K}_0$, $\tilde{\mathcal K}_1$, $\tilde{\mathcal
K}_2$ and $\tilde{\mathcal K}_3$ satisfy the set of equations \eqref{fulleom}.
Therefore the Lax connection is given by
\begin{equation}\label{laxktsup}
\mathcal{L}_\pm = \tilde{\mathcal K}_{0\pm} + z^{-1} \tilde{\mathcal K}_{1\pm} + z \tilde{\mathcal K} _{3\pm} + z^{\pm 2} \tilde{\mathcal K}_{2\pm} \ ,
\end{equation}
which in terms of $\mathcal K_{0,1,2,3}$ is
\begin{equation}\label{laxksup}
\mathcal{L}_\pm = \mathcal{K}_{0\pm} + \bar \eta \mathcal{K}_{2\pm}
+ \widehat \eta z^{\pm 2}{\mathcal{K}}_{2\pm}
+ \widehat \eta^{\frac12} \big[z^{-1}(\eta_+\mathcal{K}_{1\pm} - \eta_- \mathcal{K}_{3\pm})
+ z(\eta_+\mathcal{K}_{3\pm} - \eta_- {\mathcal{K}}_{1\pm})\big] \ .
\end{equation}
One can then also construct the Lax connection in terms of the original
currents $\mathcal J_{0,1,2,3}$. The $\mathbb{Z}_4$ symmetry \eqref{z4lax} is
generically broken. It is only present in the case that $\eta_{_L} = \pm
\eta_{_R}$, which corresponds to the deformation considered in
\cite{Delduc:2013qra,Delduc:2014kha}. The breaking of this symmetry while being
able to preserve the classical integrability of the model appears to be
intimately connected with the direct product structure of the symmetry group.

Again, as for the bosonic construction in section \ref{secbyb}, the necessity
of starting from a supercoset of the form \eqref{supercosets} is clear from
\eqref{shiftsfull}. This structure allowed us to write the full set of
equations given in \eqref{ezsup} in terms of $\mathcal{K}_0$ and
$\mathcal{K}_2$, which both take values in the Grassmann-even part of the
superalgebra $\hat{\mathfrak{g}}$ and $\mathcal{K}_1$ and $\mathcal{K}_3$, both
taking values in the Grassmann-odd part, with no restrictions. Consequently, we
could add and subtract $\mathcal{K}_0$ and $\mathcal{K}_2$ and also
$\mathcal{K}_1$ and $\mathcal{K}_3$ in \eqref{shiftsfull}.

\subsection[AdS\texorpdfstring{$_3 \times S^3 (\times S^3)$}{3 x S3 (x S3)} with B-field]{AdS\texorpdfstring{$\mathbf{_3 \times S^3 (\times S^3)}$}{3 x S3 (x S3)} with B-field}\label{secsmf}

The Metsaev-Tseytlin supercoset action for supercosets of the form
\eqref{supercosets} can alternatively be deformed, preserving integrability,
through the introduction of a B-field \cite{Cagnazzo:2012se}. The action is
given by
\begin{equation}\begin{split}\label{symsscmf}
\mathcal{S} = \int d^2x \; \operatorname{STr}[\EuScript{J}_{+}(P^{\text{b}}_{-}\EuScript{J}_{-})] -
2\text{b}\int d^3 x \; \epsilon^{abc} \, & \widetilde{\operatorname{STr}} [ \frac{2}{3}(P_2\EuScript{J}_{a})(P_2 \EuScript{J}_{b})(P_2 \EuScript{J}_{c})
+ [(P_1\EuScript{J}_{a}),(P_3 \EuScript{J}_{b})](P_2 \EuScript{J}_{c})] \ ,
\end{split}\end{equation}
which we have written in the form introduced in
\cite{Hoare:2013pma,Babichenko:2014yaa}. $\widetilde{\operatorname{STr}}$ is
defined as
\begin{equation}
\widetilde{\operatorname{STr}}[\EuScript A] = \widetilde{\operatorname{STr}} \begin{pmatrix} \mathcal{A} & 0 \\ 0 & \tilde{\mathcal{A}} \end{pmatrix}
= \operatorname{STr}[\mathcal{A}] - \operatorname{STr}[\tilde{\mathcal{A}}] = \operatorname{STr}[W\EuScript A] \ , \qquad W = \begin{pmatrix} \mathbf{1} & 0 \\ 0 &-\mathbf{1} \end{pmatrix} \ .
\end{equation}
and
\begin{equation}
P^{\text{b}}_{\pm} = P_2 \mp\frac{\sqrt{1-\text{b}^2}}2(P_1 - P_3) \ .
\end{equation}
A number of features of this action are the same as for the bosonic case
discussed in section \ref{secmf}. First, if the usual supertrace is used in the
WZ term it vanishes as a consequence of the $\mathbb{Z}_4$ automorphism of the
algebra. Second, the action \eqref{symsscmf} still has the $F_0$ gauge symmetry
defined in \eqref{gaugesup}. Third, unlike the two-parameter deformation
discussed in section \ref{secsud}, the presence of the B-field does not break
the global $\widehat F = \widehat G \times \widehat G$ symmetry of the
undeformed model.

The equation of motion following from \eqref{symsscmf} is
\begin{equation}\begin{split}
\partial_+(P_-^{\text{b}} \EuScript J_-) + &[\EuScript J_+, P_+^{\text{b}}\EuScript J_-]
+ \partial_-(P_-^{\text{b}} \EuScript J_+) + [\EuScript J_-, P_+^{\text{b}}\EuScript J_+]
\\ - \text{b}W\big(&2[P_2\EuScript J_-,P_2\EuScript J_+]
+ [P_1\EuScript J_-,P_3\EuScript J_+] + [P_3\EuScript J_-,P_1\EuScript J_+]
\\ & + [P_3\EuScript J_-,P_2\EuScript J_+] - [P_3\EuScript J_+,P_2\EuScript J_-]
- [P_1\EuScript J_-,P_2\EuScript J_+] +[P_1\EuScript J_+,P_2\EuScript J_-]\big) = 0 \ .
\end{split}\end{equation}
Projecting this equation and the flatness equation for the left-invariant
current $\EuScript J$ \eqref{z} onto $\mathfrak{f}_0$, $\mathfrak{f}_1$,
$\mathfrak{f}_2$ and $\mathfrak{f}_3$ and defining
\begin{alignat}{2}\nonumber
& \tilde{\mathcal{K}}_{0\pm} = \mathcal{J}_{0\pm} \pm \text{b} \mathcal{J}_{2\pm} \ , & \qquad &
\tilde{\mathcal{K}}_{1\pm} =\widehat{\text{b}}^\frac12(\text{b}_+\mathcal{J}_{1\pm} + \text{b}_- \mathcal{J}_{3\pm}) \ ,
\\
& \tilde{\mathcal{K}}_{2\pm} = \widehat{\text{b}} \mathcal{J}_{2\pm} \ , & \qquad &
\tilde{\mathcal{K}}_{3\pm} = \widehat{\text{b}}^\frac12(\text{b}_+\mathcal{J}_{3\pm} - \text{b}_- \mathcal{J}_{1\pm}) \ ,
\end{alignat}
with
\begin{equation}
\widehat{\text{b}} = \sqrt{1-\text{b}^2} \ , \qquad
\text{b}_+ = \sqrt{\frac{1 + \sqrt{1-\text{b}^2}}{2}} \ , \qquad
\text{b}_- = \operatorname{sign}(\text{b})\,\sqrt{\frac{1 - \sqrt{1-\text{b}^2}}{2}} \ ,
\end{equation}
we find that $\tilde{\mathcal K}_0$, $\tilde{\mathcal K}_1$,
$\tilde{\mathcal{K}}_2$ and $\tilde{\mathcal K}_3$ satisfy the equations given
in \eqref{fulleom}, and hence the Lax connection is given by
\begin{equation}
\mathcal{L}_\pm = \tilde{\mathcal K}_{0\pm} + z^{-1} \tilde{\mathcal K}_{1\pm} + z \tilde{\mathcal K}_{3\pm} + z^{\pm 2} \tilde{\mathcal K}_{2\pm} \ .
\end{equation}
In terms of the original currents $\mathcal J_{0,1,2,3}$ the Lax connection is
\begin{equation}\label{laxmixedsup}
\mathcal{L}_\pm =
\mathcal{J}_{0\pm} \pm \text{b} \mathcal{J}_{2\pm}
+ \widehat{\text{b}} z^{\pm 2}{\mathcal{J}}_{2\pm}
+ \widehat{\text{b}}^{\frac12}\big[z^{-1}(\text{b}_+\mathcal{J}_{1\pm} + \text{b}_- \mathcal{J}_{3\pm})
+ z(\text{b}_+\mathcal{J}_{3\pm} - \text{b}_- {\mathcal{J}}_{1\pm})\big] \ .
\end{equation}
As for the two-parameter deformation discussed in section \ref{secsud}, the
presence of the B-field breaks the $\mathbb{Z}_4$ symmetry \eqref{z4lax}
\cite{Cagnazzo:2012se}.

Finally let us comment that the form of the two Lax connections,
\eqref{laxksup} and \eqref{laxmixedsup}, suggests that it may be possible to
incorporate the two deformations into a three-parameter deformed model
preserving integrability.

\subsection{Comments on string theory, Virasoro constraints and \texorpdfstring{$\kappa$}{kappa}-symmetry}\label{stvcks}

The spaces AdS$_3 \times S^3$ and AdS$_3 \times S^3 \times S^3$ can be extended
to solutions of Type II supergravity in ten dimensions with the required extra
directions given by $T^4$ and $S^1$ respectively
\cite{Giveon:1998ns,Pesando:1998wm,Elitzur:1998mm}. The relation between the
Green-Schwarz string in these backgrounds and the supercoset sigma model
\eqref{mt} discussed at the beginning of this section was clarified in
\cite{Babichenko:2009dk}. In particular, the $\kappa$-symmetry of the
Green-Schwarz string can be completely fixed to give \eqref{mt} along with the
flat directions. While the resulting supercoset sigma model on its own has
eight $\kappa$-symmetries, these are broken by the coupling to the flat
directions through the Virasoro constraints, or equivalently the worldsheet
metric. As the complete supergravity backgrounds are not simple
\cite{Lunin:2014tsa} we will leave the study of $\kappa$-symmetry of the
deformed model \eqref{symbybcoset} for future work. Rather we will restrict
ourselves to outlining how the worldsheet metric should be restored in the
supercoset actions and the derivation of the corresponding contribution to the
Virasoro constraints.

The construction of the Lax connection in the earlier parts of this section was
in conformal gauge. To derive the Virasoro constraints we need to
restore the worldsheet metric $h_{\alpha\beta}$ in the actions \eqref{mt} and
\eqref{symbybcoset}. In the following we will work with the Weyl-invariant
combination of the worldsheet metric $\gamma_{\alpha\beta} = \sqrt{-h^{-1}}
h_{\alpha\beta}$ and its inverse $\gamma^{\alpha\beta} = \sqrt{-h}
h^{\alpha\beta}$. In particular, worldsheet indices will be raised and lowered
with these tensor densities. Let us recall that $\gamma_{\alpha\beta}$ and its
inverse are then symmetric and, understood as matrices, have determinant equal
to minus one. One suggestive way to restore the worldsheet metric is to
consider the following projections
\begin{equation}\label{xiproj}
(\Xi_\pm \Lambda)_\alpha = \frac12 \gamma_{\alpha\beta} (\gamma^{\beta\gamma} \mp \epsilon^{\beta\gamma}) \Lambda_\gamma \ ,
\end{equation}
where $\epsilon^{\alpha\beta}$ is the antisymmetric tensor with $\epsilon^{01}
= - \epsilon^{10} = 1$. It then follows that in conformal gauge
\begin{equation}\label{confgaugemet}
\gamma^{\alpha\beta}=\eta^{\alpha\beta} \ , \qquad \eta^{00} = -\eta^{11} = -1 \ , \quad \eta^{01} = \eta^{10} = 0 \ ,
\end{equation}
we have
\begin{equation}
(\Xi_\pm\Lambda)_\pm = \Lambda_\pm \ , \qquad (\Xi_\pm\Lambda)_\mp = 0\ ,
\end{equation}
where the light-cone coordinates are defined in footnote \ref{tension}. It is
useful to note the following set of equalities
\begin{equation}\label{general}
(\gamma^{\alpha\beta} + \epsilon^{\alpha\beta}) \Lambda_{\alpha} \tilde \Lambda_{\beta}
= 2\gamma^{\alpha\beta}(\Xi_+\Lambda)_{\alpha}(\Xi_-\tilde\Lambda)_\beta
= 2\gamma^{\alpha\beta}(\Xi_+\Lambda)_\alpha \Lambda_\beta
= 2\gamma^{\alpha\beta} \Lambda_\alpha (\Xi_-\Lambda)_\beta \ ,
\end{equation}
along with the fact that in conformal gauge these expressions reduce to
\begin{equation}\label{generalconfgauge}
- \Lambda_+ \tilde \Lambda_- \ ,
\end{equation}
where we have used the conformal-gauge metric \eqref{confgaugemet} in
light-cone coordinates
\begin{equation}
\eta^{+-} = \eta^{-+} = - \frac12 \ , \qquad \eta^{++} = \eta^{--} = 0 \ .
\end{equation}

The worldsheet metric is then restored to the supercoset action \eqref{mt} in
the following manner \cite{Delduc:2013qra}
\begin{equation}\label{mtgen}
\mathcal{S} = - \int d^2 x \; (\gamma^{\alpha\beta} + \epsilon^{\alpha\beta})\operatorname{STr}[\EuScript J_\alpha (P_- \EuScript J_\beta)]
= -2 \int d^2 x\; \gamma^{\alpha\beta}\operatorname{STr}[ \EuScript J_\alpha (P_- (\Xi_-\EuScript J)_\beta)] \ .
\end{equation}
Using \eqref{generalconfgauge} it is clear that in conformal gauge
\eqref{mtgen} indeed simplifies to \eqref{mt}. Following the same procedure for
the deformed action
\eqref{symbybcoset} we find
\begin{equation}\begin{split}\label{symbybcosetgen}
\mathcal{S} & = - \int d^2 x \;(\gamma^{\alpha\beta} + \epsilon^{\alpha\beta}) \operatorname{STr}[\EuScript J_\alpha\big( P^{\eta_{_{L,R}}}_-\frac1{1-I_{\eta_{_{L,R}}} R_f P^{\eta_{_{L,R}}}_-} \EuScript J_\beta\big)]
\\ & = - 2\int d^2 x \; \gamma^{\alpha\beta} \operatorname{STr}[\EuScript J_\alpha \big( P^{\eta_{_{L,R}}}_-\frac1{1-I_{\eta_{_{L,R}}} R_f P^{\eta_{_{L,R}}}_-} (\Xi_-\EuScript J)_\beta\big) ] \ .
\end{split}\end{equation}
Using the identities \eqref{general} one can then see that the construction of
the Lax connection in section \ref{secsud} can be naturally generalized from
conformal gauge. To do so one should replace
\begin{equation}\label{prescrip}
\partial_\pm (\mathcal{O} \EuScript J_\mp) \to -\frac12 \partial_\alpha (\mathcal{O} (\Xi_\mp\EuScript J)^\alpha) \ , \qquad
\mathcal{O}_1 \EuScript J_\pm \mathcal{O}_2 \EuScript J_\mp \to -\frac12 \mathcal{O}_1(\Xi_\pm\EuScript J)_\alpha \mathcal{O}_2(\Xi_\mp\EuScript J)^\alpha \ ,
\end{equation}
in the equation of motion \eqref{e} and flatness equation \eqref{z}. Here
$\mathcal{O}$, $\mathcal{O}_{1,2}$ denote arbitrary operators acting on the
space associated to the superalgebra. This prescription then implies that the
flatness equation \eqref{z} is generalized to
\begin{equation}
\partial_\alpha (\Xi_+\EuScript J)^\alpha -
\partial_\alpha (\Xi_-\EuScript J)^\alpha +
[(\Xi_-\EuScript J)_\alpha, (\Xi_+\EuScript J)^\alpha] = 0 \ ,
\end{equation}
which, on substituting in the definitions of the projectors $\Xi_\pm$
\eqref{xiproj}, reduces to
\begin{equation}
\epsilon^{\alpha\beta}(\partial_\alpha \EuScript J_\beta + \frac12[\EuScript J_\alpha, \EuScript J_\beta]) = 0 \ ,
\end{equation}
recovering the expected expression. The Lax connection is then given by taking
the following linear combination
\begin{equation}\begin{split}
\mathcal{L}_\alpha & =(\mathcal L_+)_\alpha + (\mathcal L_-)_\alpha \ .
\\
(\mathcal{L}_\pm)_\alpha & = (\mathcal{K}_{0\pm})_\alpha + \bar \eta (\mathcal{K}_{2\pm})_\alpha
+ \widehat \eta z^{\pm 2}({\mathcal{K}}_{2\pm})_\alpha
\\ & \hspace{40pt}+ \widehat \eta^{\frac12} \big[z^{-1}(\eta_+(\mathcal{K}_{1\pm})_\alpha - \eta_- (\mathcal{K}_{3\pm})_\alpha)
+z(\eta_+(\mathcal{K}_{3\pm})_\alpha - \eta_- ({\mathcal{K}}_{1\pm})_\alpha)\big] \ ,
\end{split}\end{equation}
where
\begin{equation}
(\EuScript{K}_{\pm})_\alpha = \mathcal{O}_\pm^{-1}(\Xi_\pm \EuScript J)_\alpha \ , \qquad
\EuScript K = \begin{pmatrix} \mathcal K & 0 \\ 0 & \tilde{\mathcal K} \end{pmatrix} \ ,
\end{equation}
is the natural generalization of \eqref{326} from conformal gauge. The
conformal-gauge flatness equation for the Lax connection is then modified to
\begin{equation}
\partial_\alpha (\Xi_+\mathcal L_+)^\alpha - \partial_\alpha (\Xi_-\mathcal L_-)^\alpha +
[(\Xi_-\mathcal L_-)_\alpha, (\Xi_+\mathcal L_+)^\alpha] = 0 \ ,
\end{equation}
which, as expected, is equivalent to
\begin{equation}
\epsilon^{\alpha\beta}(\partial_\alpha\mathcal L_\beta + \frac12[\mathcal L_\alpha,\mathcal L_\beta]) = 0 \ .
\end{equation}

Varying with respect to the worldsheet metric we find the contribution of the
supercoset action to the Virasoro constraints
\begin{equation}\label{virasorodef}
\operatorname{STr}[(P_2\mathcal{O}_\pm^{-1}\EuScript J_\alpha)(P_2\mathcal{O}_\pm^{-1}\EuScript J_\beta)-\frac12\gamma_{\alpha\beta}
(P_2\mathcal{O}_\pm^{-1}\EuScript J_\gamma)(P_2 \mathcal{O}_\pm^{-1}\EuScript J^\gamma)] + \ldots = 0 \ ,
\end{equation}
which in conformal gauge simplifies to
\begin{equation}\label{virasorodefconf}
\operatorname{STr}[(P_2\mathcal{O}_\pm^{-1}\EuScript J_{\pm})(P_2\mathcal{O}_\pm^{-1}\EuScript J_{\pm})] + \ldots = 0 \ .
\end{equation}
In \eqref{virasorodef} and \eqref{virasorodefconf} the first terms originate
from the supercoset action \eqref{mtgen}, while the ellipses denote the
contribution from the additional compact directions required for a consistent
ten-dimensional string background.

It remains an open question whether the deformation of the supercoset model can
be extended to an integrable deformation of strings in AdS$_3 \times S^3 \times
T^4$ and AdS$_3 \times S^3 \times S^3 \times S^1$. Two steps are necessary to
answer this question. First, the corresponding supergravity backgrounds would
need to be constructed \cite{Lunin:2014tsa}. Second, a $\kappa$-symmetry gauge
choice would need to be found such that the Green-Schwarz action can be
reorganized into a part corresponding to the supercoset action and a part
corresponding to the additional compact directions, as was done for the
undeformed case in \cite{Babichenko:2009dk}.

\section{Metrics}\label{secmet}

In this section we will extract explicit expressions for the metrics of the
deformed $S^3$ and AdS$_3$ sigma models. For the former we use the following
parametrization of the gauge-fixed group-valued field $f \in SU(2) \times
SU(2)$\,\foot{$\sigma_I$ are the standard Pauli matrices:
\begin{equation*}
\sigma_1 = \begin{pmatrix} 0 & 1 \\ 1 & 0 \end{pmatrix} \ , \qquad
\sigma_2 = \begin{pmatrix} 0 & -i \\ i & 0 \end{pmatrix} \ , \qquad
\sigma_3 = \begin{pmatrix} 1 & 0 \\ 0 & -1 \end{pmatrix} \ .
\end{equation*}}
\begin{equation}\label{parametrization}
f = \begin{pmatrix} \exp(\frac{i\sigma_3}{2}(\phi+\varphi)) \exp(\frac{i\sigma_1}{2} \operatorname{arcsin}r) & 0 \\ 0 & \exp(\frac{i\sigma_3}{2}(\phi-\varphi)) \exp(-\frac{i\sigma_1}{2} \operatorname{arcsin}r) \end{pmatrix} \ .
\end{equation}
For the latter we note that that the construction in section \ref{secbyb} can
be analytically continued from $S^3$ to AdS$_3$, or equivalently from $SU(2)$
to $SU(1,1)$, without any obstruction. In particular all the formulae written
in terms of group- and algebra-valued fields are the same except that the
actions should all pick up a minus sign to give the correct signature for the
target space metric. This sign flip was accounted for by the supertrace in the
supercoset construction of section \ref{secsup}. Therefore, for the deformation
of AdS$_3$ we use the following parametrization of the gauge-fixed group-valued
field $f \in SU(1,1) \times SU(1,1)$
\begin{equation}\label{parametrizations}
f = \begin{pmatrix} \exp(\frac{i\sigma_3}{2}(\psi+t)) \exp(\frac{\sigma_1}2\operatorname{arcsinh}\rho ) & 0 \\ 0 & \exp(\frac{i\sigma_3}{2}(\psi-t)) \exp(-\frac{\sigma_1}{2} \operatorname{arcsinh}\rho) \end{pmatrix} \ .
\end{equation}

Substituting \eqref{parametrization} and \eqref{parametrizations} into
\eqref{su2ssc} (flipping the overall sign in the latter case) and expanding, we
find sigma models with the three-sphere target space metric
\begin{equation}\label{su2coord2}
ds_{0,0}^2 = \frac{dr^2}{1-r^2} +(1- r^2) d\varphi^2 + r^2 d\phi^2 \ ,
\end{equation}
and the three-dimensional anti-de Sitter space target space metric
\begin{equation}\label{su11coord2}
d\sigma_{0,0}^2 = \frac{d\rho^2}{1+\rho^2} -(1+\rho^2) dt^2 + \rho^2 d\psi^2 \ ,
\end{equation}
respectively. Note that the ranges of the coordinates are
\begin{equation}\begin{split}\label{rangers}
& r \in [0,1] \ , \qquad \ \, \varphi \in (-\pi,\pi] \ , \qquad \ \ \, \phi \in (-\pi,\pi] \ .
\\
& \rho \in [0,\infty) \ , \qquad t \in (-\infty,\infty) \ , \qquad \psi \in (-\pi,\pi] \ .
\end{split}\end{equation}

The analytic continuation from $S^3$ to AdS$_3$ can be implemented at the level
of the coordinates as follows:
\begin{equation}\label{acont}
r\to - i \rho\ , \qquad \varphi \to t \ , \qquad \phi \to \psi \ ,
\end{equation}
along with flipping the overall sign and modifying the ranges of the
coordinates as in \eqref{rangers}.

\

To extract the metrics of the deformed model, we also need to specify a
particular solution of the modified classical Yang-Baxter equation for the
algebras $\mathfrak{su}(2)\oplus\mathfrak{su}(2)$ in the case of $S^3$ and
$\mathfrak{su}(1,1)\oplus\mathfrak{su}(1,1)$ for AdS$_3$. The particular
choices we will consider are the restrictions of
\begin{equation}\label{solsu2}
R\begin{pmatrix} i e_I \sigma_I & 0 \\ 0 & i \tilde e_I \sigma_I \end{pmatrix} = \begin{pmatrix} i e_I r_{IJ} \sigma_J & 0 \\ 0 & -i \tilde e_I r_{IJ} \sigma_J \end{pmatrix} \ ,
\qquad r_{I3} = r_{3I} = r_{II} = 0 \ , \quad r_{12} = -r_{21} = 1 \ .
\end{equation}
to the appropriate real forms ($e_I,\tilde e_I \in \mathbb{R}$ for
$\mathfrak{su}(2)\oplus \mathfrak{su}(2)$ and $e_3,\tilde e_3 \in \mathbb{R}$,
$e_{1,2},\tilde e_{1,2} \in i \mathbb{R}$ for $\mathfrak{su}(1,1)\oplus
\mathfrak{su}(1,1)$).

Now substituting the parametrization \eqref{parametrization} into
\eqref{su2bybcoset} we find a sigma model with the following target space
metric:
\begin{equation}\begin{split}\label{metdefs3}
ds_{\varkappa_{_+},\varkappa_{_-}}^2 = \frac1{1+\varkappa_{_-}^2(1-r^2)+\varkappa_{_+}^2r^2}\big[\frac{dr^2}{1-r^2} &+ (1-r^2)(1+\varkappa_{_-}^2(1-r^2))d\varphi^2
\\ & + r^2(1+\varkappa_{_+}^2r^2)d\phi^2 + 2\varkappa_{_+}\varkappa_{_-} r^2(1-r^2)d\varphi d\phi\big] \ ,
\end{split}\end{equation}
where $\varkappa_{_\pm}$ are defined in terms of $\varkappa_{_{L,R}}$ in
\eqref{kappmlr}. Note that there is no B-field for this background as it is a
total derivative. As shown in \cite{Hoare:2014pna} this metric is that of
Fateev's two-parameter deformation of the $S^3$ sigma model
\cite{Fateev:1996ea}. It has a $U(1)^2$ isometry corresponding to shifts in
$\varphi$ and $\phi$, which is consistent with the claim of $q$-deformed
symmetry \eqref{symbos}. The scalar curvature is
\begin{equation}\label{curvs}
4 \big[1+\varkappa_{_-}^2(1-r^2)+\varkappa_{_+}^2r^2
+ \frac12 (1+\varkappa_{_-}^2)(1+\varkappa_{_+}^2)\frac{1-\varkappa_{_-}^2(1-r^2)-\varkappa_{_+}^2r^2}{1+\varkappa_{_-}^2(1-r^2)+\varkappa_{_+}^2r^2}\big] \ .
\end{equation}

Substituting the parametrization \eqref{parametrizations} into
\eqref{su2bybcoset} (and flipping the overall sign), or alternatively
analytically continuing \eqref{metdefs3} using \eqref{acont}, we find a sigma
model with the following deformed AdS$_3$ target space metric:
\begin{equation}\begin{split}\label{metdefads3}
d\sigma_{\varkappa_{_+},\varkappa_{_-}}^2 = \frac1{1+\varkappa_{_-}^2(1+\rho^2)-\varkappa_{_+}^2\rho^2}\big[\frac{d\rho^2}{1+\rho^2} &- (1+\rho^2)(1+\varkappa_{_-}^2(1+\rho^2))dt^2
\\ & + \rho^2(1-\varkappa_{_+}^2\rho^2)d\psi^2 + 2\varkappa_{_+}\varkappa_{_-} \rho^2(1+\rho^2)dt d\psi\big] \ .
\end{split}\end{equation}
As for the deformation of the three-sphere, the B-field is a total derivative
and the metric has a $U(1)^2$ isometry, which is realized by shifts in $t$ and
$\psi$. This is again consistent with the claim of $q$-deformed symmetry. The
scalar curvature of the metric \eqref{metdefads3} is
\begin{equation}\label{curvads}
- 4 \big[1+\varkappa_{_-}^2(1+\rho^2)-\varkappa_{_+}^2\rho^2
+ \frac12 (1+\varkappa_{_-}^2)(1+\varkappa_{_+}^2)\frac{1-\varkappa_{_-}^2(1+\rho^2)+\varkappa_{_+}^2\rho^2}{1+\varkappa_{_-}^2(1+\rho^2)-\varkappa_{_+}^2\rho^2}\big] \ .
\end{equation}

Both of the metrics \eqref{metdefs3} and \eqref{metdefads3} and their
corresponding scalar curvatures \eqref{curvs} and \eqref{curvads} appear to
exhibit singularities, which we will discuss in the following sections.

\

Finally, for reference, the explicit form of the $S^3$ sigma model with
B-field in terms of the coordinates \eqref{parametrization} is given by
\begin{equation}\label{su2coord2wz}
\mathcal{S} = \int d^2 x \; \big[\frac{\partial_+ r \partial_- r}{1-r^2} +(1- r^2) \partial_+ \varphi \partial_- \varphi + r^2 \partial_+ \phi \partial_- \phi + \frac{\text{b}}2(1-2r^2)(\partial_-\varphi \partial_+\phi -\partial_-\phi\partial_+\varphi)\big] \ ,
\end{equation}
while for AdS$_3$ it is
\begin{equation}\label{su11coord2wz}
\mathcal{S} = \int d^2 x \; \big[\frac{\partial_+ \rho \partial_- \rho}{1+\rho^2} -(1+ \rho^2) \partial_+ t \partial_- t + \rho^2 \partial_+ \psi \partial_- \psi - \frac{\text{b}}2(1+2\rho^2)(\partial_-t \partial_+\psi -\partial_-\psi\partial_+t)\big] \ .
\end{equation}

\subsection[Two-parameter deformation of \texorpdfstring{$S^3$}{S3}]{Two-parameter deformation of \texorpdfstring{$\mathbf{S^3}$}{S3}}\label{secmets}

We will now discuss some features of the deformed $S^3$ metric
\eqref{metdefs3}. It is interesting to note that if we consider the following
deformation of $\mathbb{R}^4$ preserving $U(1)^2$ symmetry
\begin{equation}
dS_{\varkappa_{_+},\varkappa_{_-}}^2 = \frac{1}{1+\varkappa_{_-}^2 |Z_1|^2+\varkappa_{_+}^2 |Z_2|^2}\Big[|dZ_1|^2+|dZ_2|^2+\frac14\big(i\varkappa_{_-}(Z_1dZ_1^*-Z_1^*dZ_1)+i\varkappa_{_+}(Z_2dZ_2^*-Z_2^*dZ_2)\big)^2\Big] \ .
\end{equation}
and consider the following surface
\begin{equation}
|Z_1|^2+|Z_2|^2 = 1 \ , \qquad Z_1 = \sqrt{1-r^2}\,e^{i\varphi} \ , \qquad Z_2 = r\,e^{i\phi} \ ,
\end{equation}
embedded into this space, which for $\varkappa_{_+}=\varkappa_{_-}=0$ is just
the three-sphere embedded in $\mathbb{R}^4$, we find the metric
\eqref{metdefs3}.

If we demand that the metric \eqref{metdefs3} is real, has positive-definite
signature over the whole manifold, as defined by the coordinate ranges
\eqref{rangers}, and is not singular, then this restricts us to two regions of
parameter space. The first is the real deformation
\begin{equation}\label{realdef}
\varkappa_{_\pm} \in \mathbb{R} \ ,
\end{equation}
while the second is the imaginary deformation\,\foot{Note that for
$|k_{_\pm}| = 1$ the metric diverges. However, with a suitable overall
rescaling two of the eigenvalues of the metric are non-zero while the third
vanishes, and hence the manifold degenerates and becomes effectively
two-dimensional. For $|k_{_+}|>1$, $|k_{_-}|<1$ and $|k_{_+}|<1$, $|k_{_-}|>1$
the metric has a singularity at
\begin{equation*}
r_* = \sqrt{\frac{1-k_{_-}^2}{k_{_+}^2-k_{_-}^2}} \ ,
\end{equation*}
and has signature $(+,+,+)$ for $r<r_*$ and $(-,-,+)$ for $r>r_*$.
Furthermore, the $\mathbb{Z}_2$ transformation \eqref{z2} does not map the
range $r \in [0,1]$ onto itself. For $|k_{_\pm}|\geq 1$, $|k_{_\pm}| \neq 1$
there is no singularity, however the signature of the metric is
$(-,-,+)$.}
\begin{equation}\label{imagdef}
\varkappa_{_\pm} = i k_{_\pm} \ , \qquad k_{_\pm} \in \mathbb{R}\ , \qquad |k_{_\pm}|\leq 1 \ , \qquad |k_{_\pm}| \neq 1 \ .
\end{equation}
For these two regions the metric has a $\mathbb{Z}_2$ symmetry given by
\begin{equation}\label{z2}
r \to \frac{\sqrt{1+\varkappa_{_-}^2}\sqrt{1-r^2}}{\sqrt{1+\varkappa_{_-}^2(1-r^2) + \varkappa_{_+}^2r^2}} \ , \qquad \varphi \leftrightarrow \phi \ ,
\end{equation}
for which the range $r \in [0,1]$ is mapped onto itself. The metric is also
mapped to itself under the following transformations
\begin{alignat}{2}\nonumber
& r\to\sqrt{1-r^2} \ , \qquad \qquad \varphi \leftrightarrow \phi \ , & & \varkappa_{_+} \leftrightarrow \varkappa_{_-} \ ,
\\\label{maps}
& r \to \frac{r\sqrt{1+\varkappa_{_+}^2}}{\sqrt{1+\varkappa_{_-}^2(1-r^2) + \varkappa_{_+}^2r^2}} \ , & \qquad & \varkappa_{_+} \leftrightarrow \varkappa_{_-} \ .
\end{alignat}
For the second map, one should first interchange $\varkappa_{_+}$ and
$\varkappa_{_-}$ in the metric \eqref{metdefs3} and then perform the
transformation of $r$. Note that these two maps combined give the
$\mathbb{Z}_2$ symmetry \eqref{z2}.

There are a number of limits of interest. Setting $\varkappa_{_+} =
\varkappa_{_-} = \varkappa$ gives the squashed $S^3$ metric
\cite{Cherednik:1981df}
\begin{equation}\label{sqs3}
ds_{\varkappa,\varkappa}^2 = \frac1{1+\varkappa^2}\big[\frac{dr^2}{1-r^2} + (1-r^2)(1+\varkappa^2(1-r^2))d\varphi^2
+ r^2(1+\varkappa^2r^2)d\phi^2 + 2\varkappa^2 r^2(1-r^2)d\varphi d\phi\big] \ ,
\end{equation}
while if we take $\varkappa_{_-} = 0$ we recover the metric of
\cite{Arutyunov:2013ega}
\begin{equation}\label{mets3minus0}
ds_{\varkappa,0}^2 = \frac1{1+\varkappa^2r^2}\big[\frac{dr^2}{1-r^2} + (1-r^2)d\varphi^2\big] + r^2d\phi^2 \ ,
\end{equation}
or more precisely, its consistent truncation to a deformation of $S^3$
\cite{Hoare:2014pna}. If we alternatively take $\varkappa_{_+} = 0$ we find
\begin{equation}\label{mets3plus0}
ds_{0,\varkappa}^2 = \frac1{1+\varkappa^2(1-r^2)}\big[\frac{dr^2}{1-r^2} + r^2d\phi^2\big] + (1-r^2)d\varphi^2 \ ,
\end{equation}
which is equivalent to \eqref{mets3minus0} through the coordinate
transformations
\begin{equation}\begin{split}\label{coordmp}
r \to \sqrt{1-r^2} \ , \qquad \varphi \leftrightarrow \phi \ , \qquad \qquad \text{or} \qquad \qquad r \to \frac{r\sqrt{1+\varkappa^2}}{\sqrt{1+\varkappa^2r^2}} \ .
\end{split}\end{equation}

Considering the imaginary deformation \eqref{imagdef}, we can set $k_+ = 1$ to
give
\begin{equation}
ds_{i,ik_{_-}}^2 = \frac1{1-k_{_-}^2}\big[\frac{dr^2}{(1-r^2)^2} + r^2d\tilde\phi^2\big] + d\varphi^2\ , \qquad \tilde \phi = \phi - k_{_-}\varphi \ ,
\end{equation}
the first two terms of which are the metric of the $SU(1,1)/U(1)$ gauged WZW
model. It is worth observing that setting $k_{_+} = k_{_-}$ and then taking
$k_{_-} \to 1$, the metric degenerates to that of the two-sphere
\begin{equation}
ds_{i,i}^2 \sim \frac1{1-k_{_-}^2}\big[\frac{dr^2}{1-r^2} + r^2(1-r^2)d\tilde\phi^2\big] \ , \qquad \tilde \phi = \phi - \varphi \ ,
\end{equation}
and hence this does not commute with setting $k_{_+} = 1$ and then taking
$k_{_-} \to 1$, in which case the metric degenerates to that of the
$SU(1,1)/U(1)$ gauged WZW model
\begin{equation}
ds_{i,i}^2 \sim \frac1{1-k_{_-}^2}\big[\frac{dr^2}{(1-r^2)^2} + r^2d\tilde\phi^2\big] \ , \qquad \tilde\phi = \phi - \varphi \ .
\end{equation}

\subsection[Two-parameter deformation of AdS\texorpdfstring{$_3$}{3}]{Two-parameter deformation of AdS\texorpdfstring{$\mathbf{_3}$}{3}}\label{secmeta}

As for the deformed $S^3$ metric, we can find the metric \eqref{metdefads3} as
that of a surface embedded in a deformation of $\mathbb{R}^{2,2}$ preserving
$U(1)^2$ symmetry
\begin{equation}
d\Sigma_{\varkappa_{_+},\varkappa_{_-}}^2 = \frac{1}{1+\varkappa_{_-}^2 |Y_0|^2-\varkappa_{_+}^2 |Y_1|^2}\Big[-|dY_0|^2+|dY_1|^2-\frac14\big(i\varkappa_{_-}(Y_0dY_0^*-Y_0^*dY_0)-i\varkappa_{_+}(Y_1dY_1^*-Y_1^*dY_1)\big)^2\Big] \ .
\end{equation}
If we then consider the following surface
\begin{equation}
|Y_0|^2-|Y_1|^2 = 1 \ , \qquad Y_0 = \sqrt{1+\rho^2}\,e^{it} \ , \qquad Y_1 = \rho\,e^{i\psi} \ ,
\end{equation}
which for $\varkappa_{_+}=\varkappa_{_-}=0$ is just AdS$_3$ embedded in
$\mathbb{R}^{2,2}$, we find the metric \eqref{metdefads3}.

As discussed in section \ref{secbyb} there are two regions of parameter space
of interest. For the real deformation \eqref{realdef}, when $|\varkappa_{_+}| >
|\varkappa_{_-}|$ the metric \eqref{metdefads3} has a singularity at\,\foot{The
signature of the metric is $(-,+,+)$ for both $\rho < \rho_*$ and $\rho >
\rho_*$, however, two of the eigenvalues of the metric interchange sign either
side of the singularity.}
\begin{equation}\label{singloc1}
\rho_* = \sqrt{\frac{1+\varkappa_{_-}^2}{\varkappa_{_+}^2-\varkappa_{_-}^2}} \ ,
\end{equation}
while for $|\varkappa_{_+}| \leq |\varkappa_{_-}|$ the metric is well-defined
with signature $(-,+,+)$ for all $\rho\in[0,\infty)$. For the imaginary
deformation \eqref{imagdef}, when $1\geq|k_{_-}|>|k_{_+}|$ the metric
\eqref{metdefads3} has a singularity at\,\foot{In this case, for $\rho <
\rho_*$ the signature of the metric is $(-,+,+)$, while for $\rho > \rho_*$ it
is $(-,-,-)$.}$^{,}$\foot{If $|k_{_\pm}| = 1$ the metric diverges. However,
with a suitable overall rescaling two of the eigenvalues of the metric are
non-zero, while the third vanishes. Therefore, as for the deformation of the
three-sphere, the manifold degenerates and becomes effectively two-dimensional.
For $|k_{_+}|>1$, $|k_{_-}|\leq1$ the metric has no singularity and signature
$(-,+,+)$, while for $|k_{_+}|\leq|k_{_-}|$, $|k_{_-}|>1$ it again has no
singularity, but has signature $(-,-,-)$. For $|k_{_+}|>|k_{_-}|>1$ the metric
has a singularity at
\begin{equation*}
\rho_* = \sqrt{\frac{1+\varkappa_{_-}^2}{\varkappa_{_+}^2-\varkappa_{_-}^2}} \ ,
\end{equation*}
and has signature $(-,-,-)$ for $\rho < \rho_*$ and $(-,+,+)$ for $\rho >
\rho_*$.}
\begin{equation}\label{singloc2}
\rho_* = \sqrt{\frac{1-k_{_-}^2}{k_{_-}^2-k_{_+}^2}} \ ,
\end{equation}
while for $1\geq|k_{_+}|\geq |k_{_-}|$, $|k_{_\pm}|\neq 1$ the metric is again
well-defined with signature $(-,+,+)$ for all $\rho \in [0,\infty)$.

From the curvature \eqref{curvads} it is apparent that these singularities are
curvature singularities. Furthermore, even in the cases for which there is no
singularity at finite $\rho$, there is a singularity at $\rho \to \infty$, so
long as $\varkappa_{_+}^2 \neq \varkappa_{_-}^2$. The case $\varkappa_{_+}^2 =
\varkappa_{_-}^2$ corresponds to a special limit, which is the analytic
continuation of the squashed $S^3$ metric \eqref{sqs3}, otherwise known as
warped AdS$_3$. It therefore follows that, so long as $\varkappa_{_+}^2 \neq
\varkappa_{_-}^2$, the metric \eqref{metdefads3} has a curvature singularity
for some value of $\rho \in [0,\infty) \cup \infty$ at a finite proper
distance. It is not fully understood how to treat this singularity, which
occurs also in the deformations of the AdS$_5$ metric
\cite{Arutyunov:2013ega,Delduc:2014kha}. Therefore, in what follows we will
restrict the range of $\rho$ to $[0,\rho_*)$ where $\rho_*$ is the location of
the singularity with smallest $\rho$. We will refer to this region as the inner
region. This restriction is motivated by the fact that, for the two regions of
parameter space \eqref{realdef} and \eqref{imagdef}, this is the range of
$\rho$ for which the metric has signature $(-,+,+)$ and the isometric
coordinate $t$ plays the role of a time-like direction.

The analytic continuations of the $\mathbb{Z}_2$ transformation \eqref{z2}
and the first map in \eqref{maps} do not give corresponding relations for the
deformed AdS$_3$ metric as the range $[0,\rho_*)$ is not mapped into the
positive real numbers. The second map in \eqref{maps} does transfer over to
give
\begin{alignat}{2}\label{maps2}
\rho \to \frac{\rho\sqrt{1+\varkappa_{_-}^2}}{\sqrt{1+\varkappa_{_-}^2(1+\rho^2) - \varkappa_{_+}^2\rho^2}} \ , & \qquad & \varkappa_{_+} \leftrightarrow \varkappa_{_-} \ .
\end{alignat}

Let us briefly mention the analogues of the limits that were considered in the
deformed $S^3$ case. Setting $\varkappa_{_+} = \varkappa_{_-} = \varkappa$
gives the warped AdS$_3$ metric
\begin{equation}
d\sigma_{\varkappa,\varkappa}^2 = \frac1{1+\varkappa^2}\big[\frac{d\rho^2}{1+\rho^2} - (1+\rho^2)(1+\varkappa^2(1+\rho^2))dt^2
+ \rho^2(1-\varkappa^2\rho^2)d\psi^2 + 2\varkappa^2 \rho^2(1+\rho^2)dt d\psi\big] \ ,
\end{equation}
while if we take $\varkappa_{_-} = 0$ we recover the metric of
\cite{Arutyunov:2013ega}
\begin{equation}\label{metads3minus0}
d\sigma_{\varkappa,0}^2 = \frac1{1-\varkappa^2\rho^2}\big[\frac{d\rho^2}{1+\rho^2} - (1+\rho^2)dt^2\big] + \rho^2d\psi^2 \ ,
\end{equation}
or more precisely, its consistent truncation to a deformation of AdS$_3$
\cite{Hoare:2014pna}. If we alternatively take $\varkappa_{_+} = 0$ we find
\begin{equation}\label{metads3plus0}
d\sigma_{0,\varkappa}^2= -(1+\rho^2)dt^2 + \frac1{1+\varkappa^2(1+\rho^2)}\big[\frac{d\rho^2}{1+\rho^2} + \rho^2d\psi^2\big] \ .
\end{equation}
The first of these metrics \eqref{metads3minus0} has curvature singularities at
$\rho = \varkappa^{-1}$ and $\rho \to \infty$, while \eqref{metads3plus0} only
has one at $\rho \to \infty$. The coordinate transformation
\begin{equation}
\rho \to \frac{\rho \sqrt{1+\varkappa^2}}{\sqrt{1-\varkappa^2\rho^2}} \ ,
\end{equation}
maps the inner region of the metric \eqref{metads3plus0} ($\rho \in
[0,\infty)$) to the inner region of the metric \eqref{metads3minus0} ($\rho \in
[0,\varkappa^{-1})$). Therefore, restricting to the inner regions, the metrics
\eqref{metads3minus0} and \eqref{metads3plus0} are diffeomorphic, in analogy to
the deformation of $S^3$ discussed in section \ref{secbyb}.

It is interesting to note that the limits $\varkappa_{_+} = 0$ and
$\varkappa_{_-} = 0$ both fall into the class of models constructed in
\cite{Delduc:2013fga,Delduc:2013qra}. This is a consequence of the fact that
the modified classical Yang-Baxter equation \eqref{mcybe} is even in $R$ and
hence one can choose the relative sign of the upper left and lower right blocks
of \eqref{solsu2} to be minus (which corresponds to $\varkappa_{_-} = 0$) or
plus (corresponding to $\varkappa_{_+}=0$). The two parameter deformation
therefore encompasses both choices. In \cite{Delduc:2013fga,Delduc:2014kha} it
was shown that for compact groups these models should then be equivalent, and
indeed this is evidenced by the fact that \eqref{mets3minus0} and
\eqref{mets3plus0} are related by a coordinate redefinition. Due to the
presence of singularities the story for non-compact groups is more subtle.
However, as we have seen, the inner regions of the two possible deformations of
AdS$_3$ \eqref{metads3minus0} and \eqref{metads3plus0} are related by a
coordinate transformation. It was also shown in \cite{Delduc:2014kha} that when
deforming the AdS$_5$ metric there are three possibilities, and the metrics
\eqref{metads3minus0} and \eqref{metads3plus0} are the two possible consistent
truncations of these three metrics to three dimensions. It would be interesting
to see if the inner regions of the three deformations of AdS$_5$ are also
diffeomorphic.

Considering the imaginary deformation \eqref{imagdef}, we can set $k_{_+} = 1$
to give
\begin{equation}
d\sigma_{i,ik_{_-}}^2 = \frac1{1-k_{_-}^2}\big[\frac{d\rho^2}{(1+\rho^2)^2} + \rho^2 d\tilde\psi^2\big] - dt^2 \ , \qquad \tilde\psi = \psi - k_{_-}t \ ,
\end{equation}
the first two terms of which are the metric of the $SU(2)/U(1)$ gauged WZW
model. Note that setting $k_{_+} = k_{_-}$ and then taking $k_{_-} \to 1$, the
metric degenerates to that of $H^2$ or Euclidean AdS$_2$
\begin{equation}
d\sigma_{i,i}^2 \sim \frac1{1-k_{_-}^2}\big[\frac{d\rho^2}{1+\rho^2} + \rho^2(1+\rho^2)d\tilde\psi^2\big] \ , \qquad \tilde\psi = \psi - t \ ,
\end{equation}
and hence this does not commute with setting $k_{_+} = 1$ and then taking
$k_{_-} \to 1$, in which case the metric degenerates to that of the
$SU(2)/U(1)$ gauged WZW model
\begin{equation}
d\sigma_{i,i}^2 \sim \frac1{1-k_{_-}^2}\big[\frac{d\rho^2}{(1+\rho^2)^2} + \rho^2 d\tilde\psi^2\big] \ , \qquad \tilde\psi = \psi - t \ .
\end{equation}

In this section and section \ref{secbyb} we have considered limits in which we
do not rescale the coordinates. If we also allow rescalings then there are
number of other options, including taking $\varkappa_{_+}\to \infty$, which is
related to the mirror model and the spaces $dS_3$ and $H^3$
\cite{Hoare:2014pna,Arutyunov:2014ota,Arutyunov:2014cra}. Alternatively,
considering the direct product of the deformed spaces, a twisting can be
introduced in the $k_{_+} \to 1$ limit to keep subleading terms and give a
pp-wave type background, whose light-cone gauge-fixing \cite{Hoare:2014pna}
gives the Pohlmeyer-reduced theory for strings moving on AdS$_3 \times S^3$
\cite{Grigoriev:2008jq}.

\subsection{Near-BMN expansion}

Let us consider the sigma model with metric
$d\sigma_{\varkappa_{_+},\varkappa_{_-}}^2 + \,
ds_{\varkappa_{_+},\varkappa_{_-}}^2$, as defined in \eqref{metdefads3} and
\eqref{metdefs3} respectively, and consider fluctuations above the BMN vacuum
\cite{Berenstein:2002jq}
\begin{equation}
t = \varphi = x^0 \ .
\end{equation}
Defining
\begin{equation}
y_1 = \rho \cos \psi \ , \quad y_2 = \rho \sin \psi \ , \qquad
z_1 = r \cos \phi \ , \quad z_2 = r \sin \phi \ ,
\end{equation}
and expanding to quadratic order in $y_i$ and $z_i$ we find
\begin{equation}\begin{split}\label{quadact1}
\mathcal{S} = \frac{1}{1+\varkappa_{_-}^2} \int d^2 x \; [ & (\partial_+ y_i -\varkappa_{_+}\varkappa_{_-} \epsilon_{ij} y_j)(\partial_- y_i - \varkappa_{_+}\varkappa_{_-}\epsilon_{ik} y_k)
-(1+\varkappa_{_+}^2)(1 +\varkappa_{_-}^2)y_iy_i
\\
& + (\partial_+ z_i -\varkappa_{_+}\varkappa_{_-} \epsilon_{ij} z_j)(\partial_- z_i - \varkappa_{_+}\varkappa_{_-}\epsilon_{ik} z_k)
- (1+\varkappa_{_+}^2)(1 +\varkappa_{_-}^2)z_iz_i] \ .
\end{split}\end{equation}
Further rewriting in terms of
\begin{equation}
y = y_1 + i y_2 \ , \qquad z = z_1 + i z_2 \ ,
\end{equation}
gives
\begin{equation}\begin{split}\label{quadact2}
\mathcal{S} = \frac{1}{1+\varkappa_{_-}^2} \int d^2 x \;[ & (\partial_+ y + i \varkappa_{_+}\varkappa_{_-} y)(\partial_- y^* - i \varkappa_{_+}\varkappa_{_-} y^*)
-(1+\varkappa_{_+}^2)(1 +\varkappa_{_-}^2)y y^*
\\
& + (\partial_+ z + i \varkappa_{_+}\varkappa_{_-} z)(\partial_- z^* - i \varkappa_{_+}\varkappa_{_-} z^*)
- (1+\varkappa_{_+}^2)(1 +\varkappa_{_-}^2)z z^*] \ .
\end{split}\end{equation}
This Lagrangian describes two particles and their antiparticles with the
following dispersion relation
\begin{equation}\label{kapdp}
(e \pm \varkappa_{_+}\varkappa_{_-})^2 - p^2 - (1+\varkappa_{_+}^2)(1+\varkappa_{_-}^2) = 0 \ .
\end{equation}
Therefore they have mass $\sqrt{1+\varkappa_{_+}^2}\sqrt{1+\varkappa_{_-}^2}$
and the energy is shifted by $\varkappa_{_+}\varkappa_{_-}$ in opposite
directions for the particle and antiparticle. This is consistent with the
quadratic actions \eqref{quadact1} and \eqref{quadact2}, which are parity
invariant, and invariant under the combination of time reversal and charge
conjugation, but not the individual transformations. Finally, we note that the
mass is greater than or equal to zero for the two regions in parameter space of
interest, given in \eqref{realdef} and \eqref{imagdef}.

It is interesting to compare again with what happens for the B-field
deformation. In that case the corresponding dispersion relation takes the form
\cite{Hoare:2013pma}
\begin{equation}\label{bdp}
e^2 - (p\pm \text{b})^2 - (1-\text{b}^2) = 0 \ ,
\end{equation}
describing a particle and antiparticle with mass $\sqrt{1-\text{b}^2}$ and
spatial momentum shifted by $\text{b}$ in opposite directions. This correlates
with the fact that the B-field deformation breaks invariance under parity and
charge conjugation, but not time reversal.

It is interesting to note that the magnitude of the energy shift in
\eqref{kapdp} and the momentum shift in \eqref{bdp} are the same as the
magnitude of the shift of $\mathcal{K}_{0\pm}$ by $\mathcal{K}_{2\pm}$ in
\eqref{laxk} and \eqref{laxksup}, and the shift of $\mathcal{J}_{0\pm}$ by
$\mathcal{J}_{2\pm}$ in \eqref{laxmixed} and \eqref{laxmixedsup} respectively.
Furthermore, the masses in \eqref{kapdp} and \eqref{bdp} are the same as the
rescalings of $\mathcal{K}_{2\pm}$ in \eqref{laxk} and \eqref{laxksup}, and
$\mathcal{J}_{2\pm}$ in \eqref{laxmixed} and \eqref{laxmixedsup} respectively.

\section{R-matrices}\label{secsm}

In this section we discuss the two-parameter deformation of the R-matrices
governing the scattering above the BMN string in AdS$_3 \times S^3 \times T^4$
and AdS$_3 \times S^3 \times S^3 \times S^1$ \cite{Borsato:2012ud}. These
R-matrices are fixed by invariance under $\mathfrak{u}(1)\inplus
\mathfrak{psu}(1|1)^2 \ltimes \mathfrak{u}(1) \ltimes \mathbb{R}^3$, and are
combined together in various ways to build the light-cone gauge S-matrices of
the aforementioned AdS$_3 \times S^3 \times M^4$ string theories
\cite{Borsato:2012ud,Borsato:2013qpa,Borsato:2014exa,Hoare:2013ida,Hoare:2013lja,Lloyd:2014bsa}.

We will consider a two-parameter $q$-deformation of this algebra, conjecturing
that the associated R-matrices will underlie the light-cone gauge S-matrices
for the backgrounds constructed in sections \ref{secsup} and \ref{secmet} on
completion to full supergravity solutions \cite{Lunin:2014tsa}. Interestingly,
it transpires that, as in this section we are considering the smaller near-BMN
algebra, only one of the $q$-deformations is a genuine deformation of the
algebra, with the other parameter appearing in the representation.

\subsection[\texorpdfstring{$q$}{q}-deformed R-matrix]{\texorpdfstring{$\mathbf{q}$}{q}-deformed R-matrix}\label{symsmat}

Let us start by constructing the fundamental R-matrices for
$\mathcal{U}_q(\mathfrak{u}(1)\inplus \mathfrak{psu}(1|1)^2 \ltimes
\mathfrak{u}(1) \ltimes \mathbb{R}^3)$. The commutation relations for the
algebra $\mathfrak{u}(1)\inplus \mathfrak{psu}(1|1)^2 \ltimes \mathfrak{u}(1)
\ltimes \mathbb{R}^3$ are
\begin{alignat}{2}\nonumber
& [\mathfrak{B},\mathfrak{Q}_\pm] = \pm 2 i \mathfrak{Q}_\pm \ , & \qquad &
[\mathfrak{B},\mathfrak{S}_\pm] = \pm 2 i \mathfrak{S}_\pm \ ,
\\\nonumber
& \{\mathfrak{Q}_+,\mathfrak{S}_-\} = \mathfrak{C} + \mathfrak{M} \equiv \mathfrak{C}_{_L} \ , & \qquad &
\{\mathfrak{Q}_-,\mathfrak{S}_+\} = \mathfrak{C} - \mathfrak{M} \equiv \mathfrak{C}_{_R} \ ,
\\\label{algebra}
& \{\mathfrak{Q}_+,\mathfrak{Q}_-\} = \mathfrak{P} \ , & \qquad & \{\mathfrak{S}_+,\mathfrak{S}_-\} = \mathfrak{K} \ .
\end{alignat}
where $\mathfrak{B}$ is the $\mathfrak{u}(1)$ outer automorphism,
$\mathfrak{Q}_\pm$ and $\mathfrak{S}_\pm$ are the supercharges and
$\mathfrak{M}$, $\mathfrak{C}$, $\mathfrak{P}$ and $\mathfrak{K}$ are the
central elements. The $q$-deformation is then rather simple and amounts to the
following modification:
\begin{equation}\begin{split}\label{qalg}
\{\mathfrak{Q}_+,\mathfrak{S}_-\} & = [\mathfrak{C}_{_L}]_q = \frac{\mathfrak{V}_{_L} - \mathfrak{V}_{_L}^{-1}}{q-q^{-1}} \ , \qquad
\mathfrak{V}_{_L} \equiv q^\mathfrak{C_{_L}} \ ,
\\
\{\mathfrak{Q}_-,\mathfrak{S}_+\} & = [\mathfrak{C}_{_R}]_q = \frac{\mathfrak{V}_{_R} - \mathfrak{V}_{_R}^{-1}}{q-q^{-1}} \ , \qquad
\mathfrak{V}_{_R} \equiv q^\mathfrak{C_{_R}} \ .
\end{split}\end{equation}
The coproducts, which define the action of the generators on tensor product
representations, are deformed in the expected way
\cite{Beisert:2008tw,Hoare:2011fj} ($\mathfrak{B}$ and $\mathfrak{C}_{_{L,R}}$
have trivial coproducts)
\begin{alignat}{2}\nonumber
& \Delta(\mathfrak{Q}_+) = \mathfrak{Q}_\pm \otimes \mathbf{1} + \mathfrak{U}\mathfrak{V}_{_L}\otimes \mathfrak{Q}_+ \ , \qquad &
& \Delta(\mathfrak{Q}_-) = \mathfrak{Q}_- \otimes \mathbf{1} + \mathfrak{U}\mathfrak{V}_{_R}\otimes \mathfrak{Q}_- \ ,
\\\nonumber
& \Delta(\mathfrak{S}_+) = \mathfrak{S}_\pm \otimes \mathfrak{V}_{_R}^{-1} + \mathfrak{U}^{-1}\otimes \mathfrak{S}_\pm \ , \qquad &
& \Delta(\mathfrak{S}_-) = \mathfrak{S}_\pm \otimes \mathfrak{V}_{_L}^{-1} + \mathfrak{U}^{-1}\otimes \mathfrak{S}_\pm \ ,
\\\label{coproduct}
& \Delta(\mathfrak{P}) = \mathfrak{P} \otimes \mathbf{1} + \mathfrak{U}^2\mathfrak{V}_{_L}\mathfrak{V}_{_R}\otimes \mathfrak{P} \ , \qquad &
& \Delta(\mathfrak{K}) = \mathfrak{K} \otimes \mathfrak{V}_{_L}^{-1}\mathfrak{V}_{_R}^{-1} + \mathfrak{U}^{-2}\otimes \mathfrak{K} \ .
\end{alignat}
Following \cite{Beisert:2008tw} we have introduced both the standard
modifications associated to the $q$-deformation ($\mathfrak{V}_{_{L,R}}$) along
with the usual braiding, represented by the abelian generator $\mathfrak{U}$.
This is done according to a $\mathbb{Z}$-grading of the algebra, whereby the
charges $-2$, $-1$, $1$ and $2$ are associated to the generators
$\mathfrak{K}$, $\mathfrak{S}$, $\mathfrak{Q}$ and $\mathfrak{P}$ respectively,
while $\mathfrak{C}$, $\mathfrak{M}$ and $\mathfrak{B}$ remain uncharged. This
braiding appears in the light-cone gauge symmetry algebras for integrable
AdS/CFT systems \cite{Gomez:2006va,Plefka:2006ze}, including the AdS$_3 \times
S^3 \times T^4$ and AdS$_3 \times S^3 \times S^3 \times S^1$ examples
\cite{Borsato:2012ud,Borsato:2013qpa,Borsato:2014exa,Hoare:2013ida,Hoare:2013lja,Lloyd:2014bsa}
and allows for the existence of a non-trivial S-matrix. Note that
$\mathfrak{V}_{_{L,R}}$ and $\mathfrak{U}$ have the standard group-like
coproduct
\begin{equation}
\Delta(\mathfrak{V}_{_{L,R}}) = \mathfrak{V}_{_{L,R}} \otimes \mathfrak{V}_{_{L,R}} \ , \qquad
\Delta(\mathfrak{U}) = \mathfrak{U} \otimes \mathfrak{U} \ .
\end{equation}
We will also need to define the opposite coproduct
\begin{equation}\label{coprodop}
\Delta^{op}(\mathfrak{J}) = \mathcal{P} \Delta(\mathfrak{J}) \ ,
\end{equation}
where $\mathcal{P}$ denotes the graded permutation of the tensor product.

For the existence of an R-matrix, the coproducts for the central elements
$\mathfrak{P}$ and $\mathfrak{K}$ should be co-commutative. This implies the
following relations\,\foot{In principle the constants of proportionality could
be taken to be different. However, it is only their product that appears in the
closure conditions and R-matrices, hence we will take them to be equal.}
\begin{equation}
\mathfrak{P} = \frac{\text{h}}{2}(1-\mathfrak{V}_{_L}\mathfrak{V}_{_R}\mathfrak{U}^2) \ , \qquad
\mathfrak{K} = \frac{\text{h}}2(\mathfrak{V}_{_L}^{-1}\mathfrak{V}_{_R}^{-1}-\mathfrak{U}^{-2}) \ .
\end{equation}

We will consider the scattering of two different two-dimensional short
representations of the algebra \eqref{algebra} with the deformation
\eqref{qalg}. The first takes the form
\begin{alignat}{2}\nonumber
&\mathfrak{B}|\phi_+\rangle = -i|\phi_+\rangle \ , && \mathfrak{B}|\psi_+\rangle = i|\psi_+\rangle \ ,
\\\nonumber
&\mathfrak{Q}_+ |\phi_+\rangle = a |\psi_+\rangle \ , && \mathfrak{Q}_- |\psi_+\rangle = b |\phi_+\rangle \ ,
\\\nonumber
&\mathfrak{S}_+ |\phi_+\rangle = c |\psi_+\rangle \ , && \mathfrak{S}_- |\psi_+\rangle = d |\phi_+\rangle \ ,
\\\label{rep1}
(\mathfrak{V}_{_L},\mathfrak{V}_{_R},\,&\mathfrak{U}) |\Phi_+\rangle = (VW,VW^{-1},U)|\Phi_+\rangle \ , & \qquad & \ \ |\Phi_+\rangle \in \{|\phi_+\rangle,|\psi_+\rangle\} \ ,
\end{alignat}
while the second is
\begin{alignat}{2}\nonumber
& \mathfrak{B}|\phi_-\rangle = i|\phi_-\rangle \ , && \mathfrak{B}|\psi_-\rangle = -i|\psi_-\rangle \ ,
\\\nonumber
& \mathfrak{Q}_- |\phi_-\rangle = a |\psi_-\rangle \ , && \mathfrak{Q}_+ |\psi_-\rangle = b |\phi_-\rangle \ ,
\\\nonumber
& \mathfrak{S}_- |\phi_-\rangle = c |\psi_-\rangle \ , && \mathfrak{S}_+ |\psi_-\rangle = d |\phi_-\rangle \ ,
\\\label{rep2}
(\mathfrak{V}_{_L},\mathfrak{V}_{_R},\,&\mathfrak{U}) |\Phi_-\rangle = (VW^{-1},VW,U)|\Phi_-\rangle \ , & \qquad & \ \ |\Phi_-\rangle \in \{|\phi_-\rangle,|\psi_-\rangle\} \ .
\end{alignat}

For both these representations the anticommutation relations for the
supercharges implies the following relations:
\begin{alignat}{2}\nonumber
& ab = \frac{\text{h}}2(1-U^2V^2) \ , & \qquad & cd = \frac{\text{h}}2(V^{-2}-U^{-2}) \ ,
\\\label{relationsabcd}
& ad = \frac{VW - V^{-1}W^{-1}}{q-q^{-1}} \ , & \qquad & bc = \frac{VW^{-1} - V^{-1}W}{q-q^{-1}} \ ,
\end{alignat}
which in turn imply the following closure condition
\begin{equation}\label{closure1}
(1 - \widehat{\xi}^2)(V - V^{-1})^2 = (W - W^{-1})^2 - \widehat{\xi}^2(U-U^{-1})^2 \ ,
\end{equation}
or equivalently
\begin{equation}\label{closure1a}
(V - V^{-1})^2 = (1-\xi^2) (W - W^{-1})^2 + \xi^2(U-U^{-1})^2 \ ,
\end{equation}
where we have introduced the couplings $\xi$ and $\widehat{\xi}$ through
\begin{equation}
\widehat{\xi} = \frac{i\xi}{\sqrt{1-\xi^2}} = \frac{\text{h}}2(q-q^{-1}) \ .
\end{equation}
Conjecturing that the definitions of the energy, momentum and mass are the same
as for the undeformed case, that is
\begin{equation}\label{cmuep}
\mathfrak{C}|\Phi_\pm\rangle = \frac{e}{2} |\Phi_\pm\rangle \ , \qquad
\mathfrak{M}|\Phi_\pm\rangle = \pm\frac{m}{2} |\Phi_\pm\rangle \ , \qquad
\mathfrak{U}|\Phi_\pm\rangle = e^{\frac{i}2 \text{p}} |\Phi_\pm\rangle \ ,
\end{equation}
we find the following relations
\begin{equation}
VW = q^{\frac12(e+m)} \ , \qquad VW^{-1} = q^{\frac12(e-m)} \ , \qquad U = e^{\frac i2 \text{p}} \ .
\end{equation}
Substituting these into the closure condition \eqref{closure1} gives
\begin{equation}\label{closure2}
(1 - \widehat{\xi}^2)(q^{\frac{e}2} - q^{-\frac{e}{2}})^2
= (q^{\frac m2} - q^{-\frac m2})^2 + 4\widehat{\xi}^2\sin^2\frac{\text{p}}2 \ ,
\end{equation}
which we interpret as the dispersion relation. In section \ref{asymsmat} we
will construct the dispersion relation for the two-parameter $q$-deformation,
of which \eqref{closure2} is a special case. Therefore, we will postpone the
discussion of how to recover the undeformed dispersion relation and the
near-BMN limit to section \ref{asymsmat}.

\

To construct the R-matrices that underlie the scattering of the representations
\eqref{rep1} and \eqref{rep2} it is convenient to introduce deformations of the
Zhukovsky variables following \cite{Beisert:2011wq,Hoare:2011wr}
\begin{equation}
U^2 = W^{-2}\frac{x^+ + \xi}{x^- + \xi} = W^2\frac{x^+}{x^-}\frac{1 + x^- \xi}{1 + x^+\xi} \ , \qquad
V^2 = W^{-2}\frac{1 + x^+\xi}{1 + x^-\xi} = W^2\frac{x^+}{x^-}\frac{x^- + \xi }{x^+ + \xi} \ .
\end{equation}
In these variables the closure condition \eqref{closure1} becomes
\begin{equation}
W^{-2}(x^+ +\frac1{x^+} + \xi + \frac1\xi) = W^2(x^- + \frac1{x^-} + \xi + \frac1\xi) \ ,
\end{equation}
while the representation parameters $a$, $b$, $c$ and $d$ are
\begin{alignat}{2}\nonumber
a & = \alpha e^{-\frac{i\pi}{4}} \sqrt{\frac{\text{h}}{2}}\gamma \ , & \qquad
b & = \alpha^{-1} e^{-\frac{i\pi}{4}} \sqrt{\frac{\text{h}}{2}}\frac{\gamma}{x^- UVW} \ ,
\\\nonumber
c & = \alpha e^{\frac{i\pi}{4}}\sqrt{1-\xi^2} \sqrt{\frac{\text{h}}{2}} \frac{W\gamma}{V(x^++\xi)} \ , & \qquad
d & = \alpha^{-1} e^{\frac{i\pi}{4}}\sqrt{1-\xi^2}\sqrt{\frac{\text{h}}{2}} \frac{\gamma}{U(1+x^+ \xi)} \ ,
\\
& & \qquad \gamma & = \sqrt{i\,UVW(x^--x^+)} \ .
\end{alignat}
Here $\alpha$ parametrizes a freedom in the set of relations
\eqref{relationsabcd}. In the $q \to 1$ limit, for which we recover the
representations relevant the light-cone gauge-fixed AdS$_3 \times S^3 \times
M^4$ superstrings it is known that $\alpha = 1$, and for convenience we will
take this value from now on.

The R-matrices are completely fixed by requiring co-commutativity with the
coproduct \eqref{coproduct}
\begin{equation}
\Delta^{op}(\mathfrak{J}) \mathbb{R} = \mathbb{R}\Delta(\mathfrak{J}) \ ,
\end{equation}
where $\Delta^{op}$ is the opposite coproduct defined in \eqref{coprodop}.
Computing the R-matrix for the scattering of two particles in the same
representation we find
\begin{alignat}{2}\nonumber
\mathbb{R}^=|\phi_\pm\phi_\pm'\rangle = & S_1^= |\phi_\pm\phi_\pm'\rangle + Q_1^= |\psi_\pm\psi_\pm'\rangle
\qquad &
\mathbb{R}^=|\psi_\pm\psi_\pm'\rangle = & S_2^= |\psi_\pm\psi_\pm'\rangle + Q_2^= |\phi_\pm\phi_\pm'\rangle
\\\nonumber
\mathbb{R}^=|\phi_\pm\psi_\pm'\rangle = & T_1^= |\phi_\pm\psi_\pm'\rangle + R_1^= |\psi_\pm\phi_\pm'\rangle
\qquad &
\mathbb{R}^=|\psi_\pm\phi_\pm'\rangle = & T_2^= |\psi_\pm\phi_\pm'\rangle + R_2^= |\phi_\pm\psi_\pm'\rangle
\end{alignat}
\begin{alignat}{3}\nonumber
S_1^= &= \frac{UVW}{U'V'W'}\frac{x^- - x'^+}{x^+ - x'^-} \ ,
\qquad & S_2^= & = 1 \ , \qquad & Q_1^= = Q_2^= & = 0 \ ,
\\\label{rmat1}
T_1^= &= \frac{1}{U'V'W'}\frac{x^+ - x'^+}{x^+ - x'^-} \ ,
\qquad & T_2^= & = UVW\frac{x^- - x'^-}{x^+ - x'^-} \ ,
\qquad & R_1^= = R_2^= & = -\frac{i}{U'V'W'}\frac{\gamma\gamma'}{x^+ - x'^-} \ ,
\end{alignat}
while computing the R-matrix for the scattering of two particles in different
representations gives
\begin{alignat}{2}\nonumber
\mathbb{R}^\parallel|\phi_\pm\phi_\mp'\rangle = & S_1^\parallel |\phi_\pm\phi_\mp'\rangle + Q_1^\parallel |\psi_\pm\psi_\mp'\rangle
\qquad &
\mathbb{R}^\parallel|\psi_\pm\psi_\mp'\rangle = & S_2^\parallel |\psi_\pm\psi_\mp'\rangle + Q_2^\parallel |\phi_\pm\phi_\mp'\rangle
\\\nonumber
\mathbb{R}^\parallel|\phi_\pm\psi_\mp'\rangle = & T_1^\parallel |\phi_\pm\psi_\mp'\rangle + R_1^\parallel |\psi_\pm\phi_\mp'\rangle
\qquad &
\mathbb{R}^\parallel|\psi_\pm\phi_\mp'\rangle = & T_2^\parallel |\psi_\pm\phi_\mp'\rangle + R_2^\parallel |\phi_\pm\psi_\mp'\rangle
\end{alignat}
\begin{alignat}{3}\nonumber
T_1^\parallel & = UVWU'V'W'\frac{1-x^-x'^-}{1-x^+x'^+} \ ,
\qquad & T_2^\parallel & = 1 \ , \qquad & R_1^\parallel = R_2^\parallel & = 0 \ ,
\\\label{rmat2}
S_1^\parallel & = U'V'W'\frac{1-x^+x'^-}{1-x^+x'^+} \ ,
\qquad & S_2^\parallel & = UVW\frac{1-x^-x'^+}{1-x^+x'^+} \ ,
\qquad & Q_1^\parallel = Q_2^\parallel & = i \frac{\gamma\gamma'}{1-x^+x'^+} \ .
\end{alignat}
Note that for invariance under the action of all the symmetries the dispersion
relation needs to be imposed.

These R-matrices possess many of the properties that are required to construct
physical S-matrices describing scattering processes in an integrable theory.
They satisfy the following braiding unitarity relations
\begin{equation}
\mathbb{R}^=_{12}\mathbb{R}^=_{21} = \mathbf{1} \ , \qquad \mathbb{R}^\parallel_{12}\mathbb{R}^\parallel_{21} = \big(UVWU'V'W'\frac{1-x^-x'^-}{1-x^+x'^+}\big)\mathbf{1} \ ,
\end{equation}
the Yang-Baxter equations
\begin{alignat}{2}\nonumber
\mathbb{R}^=_{12}\mathbb{R}^=_{13}\mathbb{R}^=_{23} &=\mathbb{R}^=_{23}\mathbb{R}^=_{13}\mathbb{R}^=_{12} \ , & \qquad
\mathbb{R}^\parallel_{12}\mathbb{R}^\parallel_{13}\mathbb{R}^=_{23} &= \mathbb{R}^=_{23}\mathbb{R}^\parallel_{13}\mathbb{R}^\parallel_{12} \ ,
\\
\mathbb{R}^\parallel_{12}\mathbb{R}^=_{13}\mathbb{R}^\parallel_{23}& = \mathbb{R}^\parallel_{23}\mathbb{R}^=_{13}\mathbb{R}^\parallel_{12} \ , & \qquad
\mathbb{R}^=_{12}\mathbb{R}^\parallel_{13}\mathbb{R}^\parallel_{23}& = \mathbb{R}^\parallel_{23}\mathbb{R}^\parallel_{13}\mathbb{R}^=_{12} \ ,
\end{alignat}
and crossing relations
\begin{equation}\begin{split}
(\mathcal{C}^{-1}\otimes \mathbf{1}) \mathbb{R}^={}^{st_1}(\frac{1}{x},x') (\mathcal{C} \otimes \mathbf{1}) \mathbb{R}^\parallel(x,x') & = UVW\big(\frac{1-x^-x'^-}{1-x^+x'^-}\big) \mathbf{1} \otimes \mathbf{1} \ ,
\\
(\mathcal{C}^{-1}\otimes \mathbf{1}) \mathbb{R}^\parallel{}^{st_1}(\frac{1}{x},x') (\mathcal{C} \otimes \mathbf{1}) \mathbb{R}^=(x,x') & = UVW\big(\frac{x^--x'^+}{x^+-x'^+}\big) \mathbf{1} \otimes \mathbf{1} \ ,
\\
(\mathbf{1} \otimes \mathcal{C}^{-1}) \mathbb{R}^={}^{st_2}(x,\frac{1}{x'}) (\mathbf{1} \otimes \mathcal{C}) \mathbb{R}^\parallel(x,x') & = UVW\big(\frac{1-x^-x'^-}{1-x^+x'^-}\big) \mathbf{1} \otimes \mathbf{1} \ ,
\\
(\mathbf{1} \otimes \mathcal{C}^{-1}) \mathbb{R}^\parallel{}^{st_2}(x,\frac{1}{x'}) (\mathbf{1} \otimes \mathcal{C}) \mathbb{R}^=(x,x') & = UVW\big(\frac{x^--x'^+}{x^+-x'^+}\big) \mathbf{1} \otimes \mathbf{1} \ ,
\end{split}\end{equation}
where ${}^{st_n}$ denotes the supertranspose in factor $n$ (see, for example,
\cite{Janik:2006dc,Beisert:2008tw}) and the charge conjugation matrix is
defined as
\begin{equation}
\mathcal{C}|\phi_\pm\rangle = |\phi_\mp\rangle \ , \qquad \mathcal{C} |\psi_\pm\rangle = i |\psi_\mp\rangle \ .
\end{equation}

Finally, let us recall that in the discussion of the metrics in section
\ref{secmet}, there were two regimes of parameter space of interest,
corresponding to real $q$ (see \eqref{symsym} and \eqref{realdef}) and $q$
being a phase (see \eqref{symsym} and \eqref{imagdef}. Motivated by this we
find that the R-matrices above are also matrix unitary
\begin{equation}\label{mu}
\mathbb{R}^={}^\dagger\mathbb{R}^= = \mathbf{1} \ , \qquad
\mathbb{R}^\parallel{}^\dagger\mathbb{R}^\parallel = \mathbf{1} \ ,
\end{equation}
and the dispersion relation invariant under
conjugation\,\foot{\label{mufoot}The dispersion relation is also invariant
under conjugation for the following reality conditions
\begin{alignat*}{4}
& \xi \in (-1,1) \ , & \qquad & \widehat \xi \in i \mathbb{R} \ , & \qquad & (V^*,W^*,U^*) = (V,W,U^{-1}) \ , & \qquad & (x^\pm)^* = - \frac{x^\mp + \xi}{1+x^\mp \xi} \ ,
\\
& \xi \in i \mathbb{R} \ , & \qquad & \widehat \xi \in (-1,1) \ , & \qquad & (V^*,W^*,U^*) = (V^{-1},W^{-1},U^{-1}) \ , & \qquad & (x^\pm)^* = - x^\mp \ ,
\end{alignat*}
however, the R-matrix \eqref{rmat2} is not matrix unitary. In particular, the
non-unitarity lies in the following block
\begin{equation*}
\begin{pmatrix} S_1^\parallel & Q_2^\parallel \\ Q_1^\parallel & S_2^\parallel \end{pmatrix} \ .
\end{equation*}}
for the following reality conditions
\begin{alignat}{4}\label{mur}
& \xi \in i \mathbb{R} \ , & \qquad & \widehat \xi \in (-1,1) \ , & \qquad & (V^*,W^*,U^*) = (V,W,U^{-1}) \ , & \qquad & (x^\pm)^* = \frac{x^\mp + \xi}{1+x^\mp \xi} \ ,
\\\label{mui}
& \xi \in (-1,1) \ , & \qquad & \widehat \xi \in i \mathbb{R} \ , & \qquad & (V^*,W^*,U^*) = (V^{-1},W^{-1},U^{-1}) \ , & \qquad & (x^\pm)^* =x^\mp \ .
\end{alignat}
The first line is equivalent to those found in the AdS$_5 \times S^5$ case
\cite{Beisert:2008tw,Arutyunov:2014ota}.

This set of relations; braiding unitarity, the Yang-Baxter equations, crossing
symmetry and matrix unitarity, strongly indicate that, with the appropriate
overall factors, the R-matrices \eqref{rmat1} and \eqref{rmat2} can be used to
construct the physical S-matrices of light-cone gauge $q$-deformed AdS$_3
\times S^3 \times M^4$ string theories. This is further supported by the
presence of a similar construction in the AdS$_5 \times S^5$ case, for which
the $q$-deformed R-matrix constructed in \cite{Beisert:2008tw,deLeeuw:2011jr}
was completed to a physical S-matrix in \cite{Hoare:2011wr} through the
derivation of the overall phase. This S-matrix was then analyzed extensively
\cite{Hoare:2012fc,Arutyunov:2012zt,Arutyunov:2012ai,Arutyunov:2014ota} and in
\cite{Arutyunov:2013ega} it was shown that its near-BMN expansion at tree level
agreed with the tree-level S-matrix found from light-cone gauge-fixing the
deformed action of \cite{Delduc:2013qra}.

\

Before we discuss the two-parameter $q$-deformation, let us briefly investigate
the $\widehat{\xi} \to \infty$ limit with $q$ fixed. This is equivalent to
taking $\text{h} \to \infty$ with $q$ fixed, which in the AdS$_5 \times S^5$
case was shown \cite{Beisert:2010kk,Hoare:2011fj,Hoare:2011nd,Hollowood:2011fq}
to have a strong connection to the two-dimensional integrable theory arising as
the Pohlmeyer reduction \cite{Pohlmeyer:1975nb} of the AdS$_5 \times S^5$
superstring \cite{Grigoriev:2007bu} when $q$ is taken to be a phase. There were
complications related to the fact that the $q$-deformed R-matrix of
\cite{Beisert:2008tw} is not matrix unitary for $q$ a phase and the tree-level
S-matrix of the Pohlmeyer-reduced theory does not satisfy the classical
Yang-Baxter equation \cite{Hoare:2009fs,Hoare:2011fj}. These were partially
resolved in \cite{Hoare:2013ysa} through the vertex-to-IRF transformation,
however, what this means at the level of the string theory is somewhat unclear.
It is worth noting that there has been some interesting recent progress on this
question. In \cite{Hollowood:2014qma} it was proposed that the IRF picture
S-matrix is related to an alternative deformation, this time of the non-abelian
T-dual of the AdS$_5 \times S^5$ superstring.

For the Pohlmeyer reduction of the AdS$_3 \times S^3$ supercoset model
\cite{Grigoriev:2008jq} there are no such issues. In \cite{Hoare:2011fj} it was
shown that the Yang-Baxter equation is satisfied to one-loop order (with the
appropriate integrability-preserving one-loop counterterms), while as we have
seen above the $q$-deformed R-matrix is unitary for $q$ a phase. Furthermore,
in \cite{Hoare:2011fj} an exact integrable relativistic S-matrix whose
underlying symmetry is $\mathcal{U}_q(\mathfrak{u}(1) \inplus
\mathfrak{psu}(1|1)^2 \ltimes \mathfrak{u}(1) \ltimes \mathbb{R}^3)$ was
constructed (including overall phases), the expansion of which agreed with the
perturbative result. It is therefore natural to expect that the underlying
relativistic R-matrices will appear as limits of the R-matrices \eqref{rmat1}
and \eqref{rmat2}. Indeed, following the AdS$_5 \times S^5$ construction
\cite{Hoare:2011fj,Hoare:2011nd,Hoare:2011wr}, and taking the $\widehat{\xi}
\to \infty$ limit as follows
\begin{equation}\label{prlimit}
x^\pm = -1 + \widehat{\xi}^{-1} W^{\pm1} e^{\theta} + \mathcal{O}(\widehat{\xi}^{-2}) \ , \qquad \widehat{\xi} \to \infty \ , \qquad W = e^{\frac{i\pi}{\text{k}}} \ ,
\end{equation}
we find the following limits of the parametrizing functions
\begin{alignat}{2}\nonumber
& S_1^= = \sinh\big(\frac{\theta-\theta'}{2} - \frac{i\pi}{\text{k}}\big)\operatorname{csch}\big(\frac{\theta-\theta'}{2} + \frac{i\pi}{\text{k}}\big) \ ,
& \qquad & Q_1^= = Q_2^= = 0 \ ,
\\\nonumber
& S_2^= = 1 \ , &&
\\\nonumber
& T_1^= = T_2^= = \sinh\big(\frac{\theta-\theta'}{2}\big)\operatorname{csch}\big(\frac{\theta-\theta'}{2} + \frac{i\pi}{\text{k}}\big) \ ,
& \qquad & R_1^= = R_2^= = - i\sin\frac{\pi}{\text{k}}\operatorname{csch}\big(\frac{\theta-\theta'}{2} + \frac{i\pi}{\text{k}}\big) \ ,
\\\nonumber\vphantom{\frac12}
& T_1^\parallel = T_2^\parallel = 1 \ , & \qquad & R_1^\parallel = R_2^\parallel = 0 \ ,
\\\nonumber
& S_1^\parallel = \operatorname{sech}\big(\frac{\theta - \theta'}{2}\big)\cosh\big(\frac{\theta-\theta'}{2} + \frac{i\pi}{\text{k}}\big) \ ,
& \qquad & Q_1^\parallel =Q_2^\parallel = i \sin \frac{\pi}{\text{k}} \operatorname{sech}\big(\frac{\theta - \theta'}{2}\big) \ ,
\\\nonumber
& S_2^\parallel = \operatorname{sech}\big(\frac{\theta - \theta'}{2}\big)\cosh\big(\frac{\theta-\theta'}{2} - \frac{i\pi}{\text{k}}\big) \ , & &
\end{alignat}
which, as claimed, precisely agree with the relativistic functions found in
\cite{Hoare:2011fj} up to overall factors.

\subsection[Two-parameter \texorpdfstring{$q$}{q}-deformation of the R-matrix]{Two-parameter \texorpdfstring{$\mathbf{q}$}{q}-deformation of the R-matrix}\label{asymsmat}

In this section we will consider a two-parameter deformation of the symmetry
algebra \eqref{algebra}. It will transpire that one of these parameters can be
absorbed in the representation, recovering the one-parameter deformation
discussed in section \ref{symsmat}. Consequently the R-matrices that follow
from symmetry considerations are again given by \eqref{rmat1} and
\eqref{rmat2}, with the additional parameter entering in the definition of
$x^\pm$ and $W$ or $U$, $V$ and $W$ in terms of the energy, spatial momentum
and mass. This is similar to what occurs for the AdS$_3 \times S^3 \times M^4$
backgrounds with a B-field \cite{Cagnazzo:2012se}, in which case the symmetry
is undeformed and the representations contain the information pertinent to the
deformation \cite{Hoare:2013pma,Hoare:2013ida,Hoare:2013lja,Lloyd:2014bsa}.

Starting again from the algebra \eqref{algebra}, the natural candidate for the
two-parameter $q$-deformation is to separately deform the central elements
$\mathfrak{C}_{_{L,R}}$ as follows:
\begin{equation}\begin{split}
& \{\mathfrak{Q}_+,\mathfrak{S}_-\} = [\mathfrak{C}_{_L}]_{q_{_L}} = \frac{\mathfrak{V}_{_L} - \mathfrak{V}_{_L}^{-1}}{q_{_L}-q_{_L}^{-1}} \ , \qquad
\mathfrak{V}_{_L} \equiv q_{_L}^{\mathfrak{C}_{_L}} \ ,
\\
& \{\mathfrak{Q}_-,\mathfrak{S}_+\} = [\mathfrak{C}_{_R}]_{q_{_R}} = \frac{\mathfrak{V}_{_R} - \mathfrak{V}_{_R}^{-1}}{q_{_R}-q_{_R}^{-1}} \ , \qquad
\mathfrak{V}_{_R} \equiv q_{_R}^{\mathfrak{C}_{_R}} \ .
\end{split}\end{equation}
Let us now define a place-holding parameter $q$ such that
\begin{equation}\label{php}
q_{_L} = q^{\rho_{_L}} \ , \qquad q_{_R} = q^{\rho_{_R}} \ .
\end{equation}
Then the rescaled generators\,\foot{Recall that $[x]_q = \frac{q^x - q^{-x}}{q
- q^{-1}}$.}
\begin{equation}\begin{split}\label{rescaled}
& \tilde{\mathfrak{Q}}_+ = \sqrt{[\rho_{_L}]_q} \, \mathfrak{Q}_+ \ , \qquad
\tilde{\mathfrak{Q}}_- = \sqrt{[\rho_{_R}]_q} \, \mathfrak{Q}_- \ , \qquad
\tilde{\mathfrak{S}}_+ = \sqrt{[\rho_{_R}]_q} \, \mathfrak{S}_+ \ , \qquad
\tilde{\mathfrak{S}}_- = \sqrt{[\rho_{_L}]_q} \, \mathfrak{S}_- \ ,
\\
& \tilde{\mathfrak{C}}_{_L} = \rho_{_L} \mathfrak{C}_L \ , \qquad \ \ \,
\tilde{\mathfrak{C}}_{_R} = \rho_{_R} \mathfrak{C}_R \ , \qquad \ \ \,
\tilde{\mathfrak{P}} = \sqrt{[\rho_{_L}]_q}\sqrt{[\rho_{_R}]_q}\mathfrak{P} \ , \qquad \ \ \,
\tilde{\mathfrak{K}} = \sqrt{[\rho_{_L}]_q}\sqrt{[\rho_{_R}]_q}\mathfrak{K} \ ,
\end{split}\end{equation}
satisfy the one-parameter $q$-deformed algebra discussed in section
\ref{symsmat}. If we then followed the derivation in section \ref{symsmat} with
the rescaled generators \eqref{rescaled} their coproducts would be given by
\eqref{coproduct} with
\begin{equation}
\mathfrak{V}_{_{L,R}} \to \tilde{\mathfrak{V}}_{_{L,R}} = q^{\tilde{\mathfrak{C}}_{_{L,R}}} \ .
\end{equation}
Observing that
\begin{equation}
\tilde{\mathfrak{V}}_{_{L,R}} = q^{\tilde{\mathfrak{C}}_{_{L,R}}} = q_{_{L,R}}^{\mathfrak{C}_{_{L,R}}} = \mathfrak{V}_{_{L,R}} \ ,
\end{equation}
we see that the coproducts for the unscaled generators in \eqref{rescaled} take
the expected form for a $q$-deformed symmetry, and hence it follows that the
new parameter can be absorbed into the representation.

Motivated by this we modify the definition of the first representation
\eqref{rep1} as follows
\begin{alignat}{2}\nonumber
& \mathfrak{B}|\phi_+\rangle = -i|\phi_+\rangle\ , & &\mathfrak{B}|\psi_+\rangle = i|\psi_+\rangle \ ,
\\\nonumber
& \mathfrak{Q}_+ |\phi_+\rangle = \frac{a}{\sqrt{[\rho_{_L}}]_q} |\psi_+\rangle \ , & &\mathfrak{Q}_- |\psi_+\rangle = \frac{b}{\sqrt{[\rho_{_R}}]_q} |\phi_+\rangle \ ,
\\\nonumber
& \mathfrak{S}_+ |\phi_+\rangle = \frac{c}{\sqrt{[\rho_{_R}}]_q} |\psi_+\rangle \ , & &\mathfrak{S}_- |\psi_+\rangle = \frac{d}{\sqrt{[\rho_{_L}}]_q} |\phi_+\rangle \ ,
\\\label{repq1}
(\mathfrak{V}_{_L},\mathfrak{V}_{_R},\,& \mathfrak{U}) |\Phi_+\rangle = (VW,VW^{-1},U)|\Phi_+\rangle\ , & \qquad & \ \ |\Phi_+\rangle \in \{|\phi_+\rangle,|\psi_+\rangle\} \ ,
\end{alignat}
and similarly for the second representation \eqref{rep2}
\begin{alignat}{2}\nonumber
& \mathfrak{B}|\phi_-\rangle = i|\phi_-\rangle \ , & & \mathfrak{B}|\psi_-\rangle = -i|\psi_-\rangle \ ,
\\\nonumber
& \mathfrak{Q}_- |\phi_-\rangle = \frac{a}{\sqrt{[\rho_{_R}}]_q} |\psi_-\rangle \ , & & \mathfrak{Q}_+ |\psi_-\rangle = \frac{b}{\sqrt{[\rho_{_L}}]_q} |\phi_-\rangle \ ,
\\\nonumber
& \mathfrak{S}_- |\phi_-\rangle = \frac{c}{\sqrt{[\rho_{_L}}]_q} |\psi_-\rangle \ , & & \mathfrak{S}_+ |\psi_-\rangle = \frac{d}{\sqrt{[\rho_{_R}}]_q} |\phi_-\rangle \ ,
\\\label{repq2}
(\mathfrak{V}_{_L},\mathfrak{V}_{_R},\,&\mathfrak{U}) |\Phi_-\rangle = (VW^{-1},VW,U)|\Phi_-\rangle \ , & \qquad & \ \ |\Phi_-\rangle \in \{|\phi_-\rangle,|\psi_-\rangle\} \ .
\end{alignat}
From here one can proceed as in section \ref{symsmat} arriving at the
R-matrices \eqref{rmat1} and \eqref{rmat2} and the closure condition
\eqref{closure1}. As outlined above, the subtlety now lies in how to define of
$x^\pm$ and $W$ or $U$, $V$ and $W$ in terms of the energy, spatial momentum
and mass.

The crucial observation is that the R-matrices \eqref{rmat1} and \eqref{rmat2}
and the closure condition \eqref{closure1} have no explicit dependence on the
place-holding parameter $q$ introduced in \eqref{php} or $\text{h}$. This can
be seen by noting that all the dependence comes through $V$, $W$, $x^\pm$ and
$\widehat{\xi}$ (or equivalently $\xi$). If we preserve the identifications
given in \eqref{cmuep} we find the following relations
\begin{equation}\label{rel1}
VW = q_{_L}^{\frac12(e+m)} \ , \qquad VW^{-1} = q_{_R}^{\frac12(e-m)} \ , \qquad U = e^{\frac i2 \text{p}} \ ,
\end{equation}
for the first representation \eqref{repq1} and
\begin{equation}\label{rel2}
VW = q_{_R}^{\frac12(e+m)} \ , \qquad VW^{-1} = q_{_L}^{\frac12(e-m)} \ , \qquad U = e^{\frac i2 \text{p}} \ ,
\end{equation}
for the second \eqref{repq2}. This demonstrates explicitly that when written in
terms of the physical kinematical variables, energy, spatial momentum and mass,
the explicit dependence of the R-matrices \eqref{rmat1} and \eqref{rmat2} and
the closure condition \eqref{closure1} will be on the parameters $q_{_L}$,
$q_{_R}$ and $\widehat{\xi}$ (or equivalently $\xi$).

This then clarifies the role of the parameter $q$ introduced in equation
\eqref{php} as purely a place holder. It also demonstrates that $\text{h}$
plays a similar role in the two-parameter deformation. Consequently the three
parameters we take as independent are $q_{_L}$, $q_{_R}$ and $\widehat{\xi}$
(or equivalently $\xi$).

Substituting the relations \eqref{rel1} and \eqref{rel2} into the closure condition \eqref{closure1} we find
\begin{equation}\begin{split}\label{ddd}
(1-\widehat{\xi}^2)(q_{_L}^{\frac14(e \pm m)}& q_{_R}^{\frac14(e \mp m)}-q_{_L}^{-\frac14(e \pm m)}q_{_R}^{-\frac14(e \mp m)})^2 =
\\
& (q_{_L}^{\frac14(e \pm m)}q_{_R}^{-\frac14(e \mp m)}-q_{_L}^{-\frac14(e \pm m)}q_{_R}^{\frac14(e \mp m)})^2 + 4\widehat{\xi}^2 \sin^2 \frac{\text{p}}2 \ ,
\end{split}\end{equation}
which we interpret as the dispersion relation of the two-parameter deformation.

\

Let us now discuss how to recover the undeformed dispersion relation in the
$q_{_{L,R}} \to 1$ limit and the near-BMN dispersion \eqref{kapdp}. If this
deformed R-matrix and closure condition do indeed underlie the light-cone gauge
S-matrices of strings in the deformed backgrounds then the three parameters
$q_{_L}$, $q_{_R}$ and $\widehat{\xi}$ should be mapped to the three parameters
of the supercoset actions in section \ref{secsud}. These were the deforming
parameters $\varkappa_{_L}$, $\varkappa_{_R}$ and the effective string tension
$h$. To relate the two sets of parameters, we start by using the semiclassical
identifications of $q_{_{L,R}}$ in terms of $\varkappa_{_{L,R}}$ given in
\eqref{symsym}
\begin{equation}\label{iiii}
q_{_{L,R}} = e^{-\frac{\varkappa_{_{L,R}}}{h}} \ .
\end{equation}
It will then transpire that to recover the expected limits we need to fix
\begin{equation}\begin{split}\label{iv}
\widehat{\xi}^2 & = \frac{\varkappa_{_L}\varkappa_{_R}}{1+\frac14(\varkappa_{_L} + \varkappa_{_R})^2} =
\frac{\varkappa_{_+}^2 - \varkappa_{_-}^2}{1+\varkappa_{_+}^2} \ ,
\qquad \xi^2 = - \frac{\varkappa_{_L}\varkappa_{_R}}{1+\frac14(\varkappa_{_L} - \varkappa_{_R})^2}
= - \frac{\varkappa_{_+}^2 - \varkappa_{_-}^2}{1+\varkappa_{_-}^2} \ ,
\end{split}\end{equation}
at least at leading order in the two expansions discussed below. Let us recall
that $\varkappa_{_\pm}$ are defined in terms of $\varkappa_{_{L,R}}$ in
\eqref{kappmlr}. Of course all of these relations may receive subleading
corrections. Note that in the case $\varkappa_{_L} = \varkappa_{_R} =
\varkappa$ we find that
\begin{equation}
\xi^2 = -\varkappa^2 \ ,
\end{equation}
which agrees with the identification found in the $q$-deformed AdS$_5 \times
S^5$ model \cite{Arutyunov:2013ega,Arutyunov:2014ota}. This is consistent since
taking $\varkappa_{_L} = \varkappa_{_R}$ corresponds to the one-parameter
deformation of \cite{Delduc:2013qra}. In particular, as discussed in section
\ref{secmeta}, this limit ($\varkappa_{_-} = 0$) gives the truncation of the
model considered in \cite{Arutyunov:2013ega}. This provides additional
motivation for the identification \eqref{iiii}, as in principle there is a
freedom in the relative sign of $\varkappa_{_{L}}$ and $\varkappa_{_R}$.
Furthermore, the relativistic Pohlmeyer limit should be given by
$\varkappa_{_+}^2 = -1$ (with $\varkappa_{_-} = 0$) \cite{Hoare:2014pna},
which, from \eqref{iv}, implies that $\widehat{\xi} \to \infty$. This is
consistent with the limit discussed in \eqref{prlimit}.

Assuming the identifications \eqref{iiii} and \eqref{iv} are exact and
requiring matrix unitarity of the R-matrices \eqref{mu} places additional
restrictions on the parameters $\varkappa_{_\pm}$. First let us recall that in
the discussion of the metrics in section \ref{secmet} there were two regimes of
interest. The first corresponds to real $q_{_{L,R}}$ (see \eqref{symsym} and
\eqref{realdef}) and hence real $V$ and $W$. From \eqref{mur} we see that this
requires $\xi \in i\mathbb{R}$, $\widehat{\xi} \in (-1,1)$, which combining
with \eqref{iv} implies that $\varkappa_{_+}^2 \geq \varkappa_{_-}^2$.
Similarly for the second regime, corresponding to $q_{_{L,R}}$ being a phase
(see \eqref{symsym} and \eqref{imagdef}), we find that $1 \geq k_{_+}^2 \geq
k_{_-}^2$, $k_{_\pm}^2 \neq 1$. It is interesting to note that these regimes
(excluding $\varkappa_{_+}^2 = \varkappa_{_-}^2$ and $k_{_+}^2 = k_{_-}^2$) are
the same as those for which the deformed AdS$_3$ metric has a singularity at
finite $\rho$. Furthermore, the location of this singularity \eqref{singloc1},
\eqref{singloc2} is related to $\xi$ in the following simple manner
\begin{equation}
\rho_* = \sqrt{-\xi^{-2}} \ .
\end{equation}

It is unclear whether the apparent non-unitarity in the complementary regimes,
$\varkappa_{_+}^2 < \varkappa_{_-}^2$ and $k_{_+}^2 < k_{_-}^2 \leq 1$, can be
remedied. Substituting into \eqref{iv} we see that they correspond to the
reality conditions discussed in footnote \ref{mufoot}, for which the dispersion
relation is invariant under conjugation, but the R-matrix \eqref{rmat2} is not
matrix unitary. It is worth noting that the ranges for which the R-matrices are
unitary are mapped onto their complements by \eqref{maps} and \eqref{maps2}.
However, this symmetry need not be preserved by the full background. It is
therefore possible that in the action the problem will manifest itself when
one considers the fermions.

Substituting \eqref{iiii} and \eqref{iv} into the dispersion relation
\eqref{ddd} gives
\begin{equation}\label{dddd}
(1+\varkappa_{_-}^2)\sinh^2\big(\frac{\varkappa_{_+} e \pm \varkappa_{_-} m}{2h}\big)
- (1+\varkappa_{_+}^2)\sinh^2\big(\frac{\varkappa_{_+} m \pm \varkappa_{_-} e}{2h}\big)
- (\varkappa_{_+}^2 - \varkappa_{_-}^2)\sin^2\frac{\text{p}}2 = 0 \ .
\end{equation}
To implement the $q_{_{L,R}}\to 1$ limit, we take $\varkappa_{_{L,R}}\to 0$, or
equivalently $\varkappa_{_\pm} \to 0$, at the same rate. The leading order term
in the expansion is at quadratic order and proportional to $\varkappa_{_+}^2 -
\varkappa_{_-}^2$. As claimed, this term gives the undeformed dispersion
relation \cite{Borsato:2012ud,Borsato:2013qpa}
\begin{equation}
e^2 = m^2 + 4h^2\sin^2\frac{\text{p}}2 \ .
\end{equation}

To take the large $h$ near-BMN expansion we introduce the near-BMN momentum
\begin{equation}
p = h\text{p} \ .
\end{equation}
The leading order term in this expansion then occurs at $\mathcal{O}(h^{-2})$.
We find that the dispersion relation \eqref{dddd} at this order is
equivalent to
\begin{equation}
(e \pm m\varkappa_{_+}\varkappa_{_-})^2 - p^2 - m^2 (1+\varkappa_{_+}^2)(1+\varkappa_{_-}^2) = 0 \ ,
\end{equation}
which, setting $m = 1$, agrees with the near-BMN dispersion relation
\eqref{kapdp} found from the expansion of the coset action.

\

To conclude this section let us make a brief comment on the possibility of
including a B-field from the perspective of the R-matrices. For the
undeformed AdS$_3 \times S^3 \times T^4$ model the addition of the B-field
does not modify the symmetry of the string background. The additional parameter
appears in the S-matrix through a deformation of the representations. In
particular, it is consistent with the coproducts for the action of the
generator $\mathfrak{M}$ to have a linear dependence on the spatial momentum
$\text{p}$ as both have a trivial coproduct \cite{Hoare:2013lja}. As in the
discussions relating to the deformation of the supercoset sigma model, this
again suggests that it may be possible to incorporate the two deformations into
a three-parameter deformed model preserving integrability.

\section{Comments}

In this article we have investigated the existence of a two-parameter
integrable deformation of strings moving in AdS$_3 \times S^3 \times T^4$ and
AdS$_3 \times S^3 \times S^3 \times S^1$, for which the global symmetry
$\widehat G \times \widehat G$ is $q$-deformed asymmetrically,
$\mathcal{U}_{q_{_L}}(\widehat{G}) \times \mathcal{U}_{q_{_R}}(\widehat G)$.
Two constructions providing evidence for such a deformation were described. The
first was a two-parameter deformation of the Metsaev-Tseytlin supercoset sigma
model for supercosets with isometry of the form $\widehat G \times \widehat G$,
generalizing the construction of \cite{Delduc:2013qra}. The second was a
two-parameter deformation of the $\mathfrak{u}(1) \inplus \mathfrak{psu}(1|1)^2
\ltimes \mathfrak{u}(1) \ltimes \mathbb{R}^3$-invariant R-matrices, which
underlie the scattering above the BMN string in these backgrounds.

\

In section \ref{secmet} the deformed supercoset sigma model was used to extract
the deformation of the metric and B-field (which in this case is a total
derivative). To fully demonstrate the existence of the two-parameter integrable
deformation of the string theories one would need to construct the full
supergravity background \cite{Lunin:2014tsa}, and find a $\kappa$-symmetry
gauge such that the corresponding Green-Schwarz action matches the deformed
supercoset sigma model. It is worth noting that the two-parameter deformation
of the AdS$_3$ metric in general has a curvature singularity at finite proper
distance. It is currently not clear how to treat this singularity -- better
understanding may come from the study of classical string solutions in the
deformed AdS$_3$ space
\cite{Kameyama:2014bua,Kameyama:2014vma,Kameyama:2014via,Frolov:talk}.

\

In section \ref{asymsmat} a two-parameter deformation of the dispersion
relation was proposed. To verify this one could study how classical strings,
for example the giant magnon, are affected by the deformation. For the
one-parameter deformation of \cite{Delduc:2013qra} such solutions were
considered in
\cite{Arutyunov:2014ota,Khouchen:2014kaa,Ahn:2014aqa,Arutyunov:2014cda,Banerjee:2014bca}.
It is also important to check, for example through direct perturbative
computations as was done in \cite{Arutyunov:2013ega} for the AdS$_5 \times S^5$
case, that the $q$-deformed R-matrices constructed in section \ref{symsmat}
indeed underlie the scattering above the BMN string.

A related open question is to derive overall phases such that these R-matrices
are matrix unitary, braiding unitary and crossing symmetric. That is, they can
be understood as physical scattering matrices. In the AdS$_5 \times S^5$ case
\cite{Hoare:2011wr} this amounted to replacing the gamma functions in the DHM
representation \cite{Dorey:2007xn} of the phase \cite{Beisert:2006ez} with
$q$-deformed gamma functions. In the AdS$_3 \times S^3$ case, a conjecture for
the undeformed phases for constant $m$ (i.e. independent of energy and spatial
momentum) was given in \cite{Borsato:2013hoa}. However, naively these proposals
do not appear to be amenable to such a simple deformation.

\

In this article we have highlighted certain key similarities between the
two-parameter $q$-deformation and the deformation of \cite{Cagnazzo:2012se} in
which the background is supported by a mix of RR and NSNS fluxes. These
comparisons suggest that there is naturally space for a three-parameter
integrable deformation.

The two-parameter metrics in section \ref{secmet} contain the squashed
three-sphere \cite{Cherednik:1981df} and warped AdS$_3$ metrics as particular
limits. In this case it is known that the usual B-field with arbitrary
coefficient can be introduced while preserving integrability. Recent progress
in extending these backgrounds to supergravity solutions
\cite{Orlando:2010ay,Orlando:2012hu} and understanding their integrable
structure
\cite{Orlando:2010yh,Kawaguchi:2011mz,Kawaguchi:2013gma,Delduc:2014uaa} suggest
that this might provide a strong starting point to find the three-parameter
deformation.

Furthermore, the two-parameter deformation of the $S^3$ sigma model
\cite{Fateev:1996ea} was generalized in \cite{Lukyanov:2012zt} to a
four-parameter deformation including a B-field. It is an open question as to
whether this can be extended to a deformation of the AdS$_3 \times S^3 \times
T^4$ and AdS$_3 \times S^3 \times S^3 \times S^1$ string backgrounds.

\

To conclude, let us note that a proposal was recently made for an integrable
deformation of the non-abelian T-dual of the AdS$_5 \times S^5$ superstring
\cite{Hollowood:2014qma}, based on the bosonic deformations of
\cite{Sfetsos:2013wia,Hollowood:2014rla}. It is claimed that this model is
related to the $q$-deformation in the case that $q$ is a phase. It would be
interesting to study this deformation for lower-dimensional AdS backgrounds
\cite{Sfetsos:2014cea}, in particular to see if a double deformation, analogous
to that considered in this article, can be implemented.

\section*{Acknowledgments}

I would like to thank S.~Frolov, R.~Roiban, A.~Tseytlin and S.~J.~van~Tongeren
for useful discussions and helpful comments, and I am grateful to
D.~Berenstein, L.~Bianchi, R.~Borsato, F.~Delduc, V.~Forini, T.~J.~Hollowood,
M.~Magro, T.~Matsumoto, J.~L.~Miramontes, D.~Schmidtt, A.~Sfondrini,
O.~Ohlsson~Sax B.~Vicedo and L.~Wulff for related discussions. I would also
like to thank A.~Tseytlin for valuable comments on the draft. This work is
funded by the DFG through the Emmy Noether Program ``Gauge Fields from
Strings'' and SFB 647 ``Space - Time - Matter. Analytic and Geometric
Structures.''

\bibliographystyle{nb}
\bibliography{ads3def}

\end{document}